\documentclass[pre,reprint,groupedaddress,nofootinbib]{revtex4-1}

\usepackage{amssymb}
\usepackage{graphicx}
\newcommand \be  {\begin{equation}}
\newcommand \bea {\begin{eqnarray} \nonumber }
\newcommand \ee  {\end{equation}}
\newcommand \eea {\end{eqnarray}}
\newcommand{\textsmall}[1] {\text{\scriptsize #1}}

\begin{document}

\title{Agent-based models for latent liquidity and concave price impact}
\author{Iacopo Mastromatteo}
\author{Bence T\'oth}
\author{Jean-Philippe Bouchaud}
\affiliation{Capital Fund Management, 23-25, Rue de l'Universit\'e 75007 Paris, France}

\begin{abstract}
We revisit the ``$\varepsilon$-intelligence'' model of T\'oth et al.~(2011), that was proposed as a minimal framework to understand the square-root dependence of the impact of meta-orders 
on volume in financial markets. The basic idea is that most of the daily liquidity is ``latent'' and furthermore vanishes linearly around the current price, as a consequence of the diffusion of the price itself. However, the numerical implementation of T\'oth et~al.~(2011) was criticised as being unrealistic, in particular because all the ``intelligence'' was conferred to market orders, while limit orders
were passive and random. In this work, we study various alternative specifications of the model, for example allowing limit orders to react to the order flow, or changing the execution protocols.  
By and large, our study lends strong support to the idea that the square-root impact law is a very generic and robust property that requires very few ingredients to be valid.
We also show that the transition from super-diffusion to sub-diffusion reported in T\'oth et al.~(2011) is in fact a cross-over, but that the original model can be slightly altered in order to give rise 
to a genuine phase transition, which is of interest on its own. We finally propose a general theoretical framework to understand how a non-linear impact may appear
even in the limit where the bias in the order flow is vanishingly small. 
\end{abstract}

\maketitle

\section{Introduction}

Understanding price impact is arguably one of the most important problems in financial economics. From a theoretical standpoint, price impact is the transmission belt that allows private information
to be reflected by prices. But by the same token, it is also the very mechanism by which prices can be distorted, or even crash, under the influence of uninformed trades and/or fire-sale deleveraging. Price impact is also 
a cost for trading firms -- in fact the dominant one when assets under management become substantial. As a rough order of magnitude, impact costs for individual stocks are ten times larger than fixed costs
for a firm that trades a mere $1 \%$ of the average daily volume. 
Now, the simplest guess is that price impact should be linear, i.e.\ proportional to the (signed) volume of a transaction. This is in fact the assumption of the seminal microstructure model proposed by 
Kyle in 1985~\cite{Kyle:1985}. This paper has had a profound influence on the field, with over 6000 citations as of mid-2013. A linear impact model is at the core of many different studies, concerning for example
optimal execution strategies, liquidity estimators, agent based models, volatility models, etc. 

Quite surprisingly, however, the last 15 years have witnessed mounting empirical evidence invalidating the
simple linear impact framework, suggesting instead a sub-linear, square-root like growth of impact with volume, often dubbed the ``square-root impact law''.\footnote{Empirical papers consistently
measuring an impact curve close to square root date back to 1997 \cite{Torre:1997}, see also \cite{Almgren:2005,Moro:2009,Toth:2011} and Refs.\ therein. More recent results again support the same law, see
\cite{Kyle:2012,Bershova:2013,Waelbroeck:2013} and Figs.~\ref{fig:DISCUSImp} and~\ref{fig:GammaVSDelta} below.} One should however carefully distinguish at this point
different definitions of ``price impact'' that lead to very different dependence in volume. For example, the average impact of a {\it single market order} is found to be a strongly concave function of the volume, with a significant 
dependence on the microstructure (tick size, order priority, etc.). The concavity is primarily the result of a conditioning effect: the size of a market order very rarely exceeds the total volume of limit orders that
sit at the opposite best price. Therefore, large market orders match large outstanding volumes, and result in small price changes. 

The square-root impact law that we have been referring to above is both more relevant and more 
universal. It concerns the average impact of a ``meta-order'' of size $Q$, which is a sequence of orders in the same direction from the same investor, incrementally executed in the market using either market or limit orders, 
that sum up to a certain quantity $Q$. This definition is more relevant because trades are usually much too large to be executed in a single shot, but must rather be fragmented in (many) small orders that are executed 
progressively. The impact of a meta-order is also surprisingly universal: the square-root law has been reported by many different groups, and seems to hold for completely different markets (equities,
futures, FX, etc.), epochs (from the mid-nineties, when liquidity was provided by market makers, to the present day electronic markets), microstructure (small ticks vs.\ large ticks), market participants and underlying
trading strategies (fundamental, technical, etc.) and style of execution (using limit or market orders -- see Fig.~\ref{fig:DISCUSImp}; with high or low participation rate, etc.). In all these cases, the {\it average} relative price difference $\mathcal I$ between the 
first and the last trade of a meta-order of volume $Q$ is well described by the following law:
\begin{equation}
  \label{eq:1}
  \mathcal I = Y \sigma_D \left( \frac Q {V_D}\right)^\delta \;,
\end{equation}
where $\delta$ is an exponent in the range 0.4 -- 0.7, $Y$ is a coefficient of order unity and $\sigma_D,V_D$ are respectively daily volatility and daily traded volume~\cite{Almgren:2005,Moro:2009,Toth:2011}, see also~\cite{Kyle:2012,Bershova:2013,Waelbroeck:2013}. The proprietary data of CFM published in~\cite{Toth:2011} corresponds to nearly 500,000 meta-orders between June 2007 and December 2010 on a variety of futures contracts, and leads to $\delta \approx 0.5$ for small tick contracts and $\delta \approx 0.6$ for large tick contracts, for $Q$ ranging from a few $10^{-4}$ to a few percent of the average daily liquidity. Newer data up to the end of 2012 leave these estimates unchanged. As an illustrative example, we plot in Fig.~\ref{fig:DISCUSImp}  the impact curve obtained for a specific futures contract (GBP), while in Fig.~\ref{fig:GammaVSDelta} we represent the impact exponents estimated on a set of representative futures contracts. CFM's data on individual stocks is also compatible with $\delta \approx 0.5-0.7$ for all geographical zones.
\begin{figure}[hbtp]
  \centering
  \includegraphics{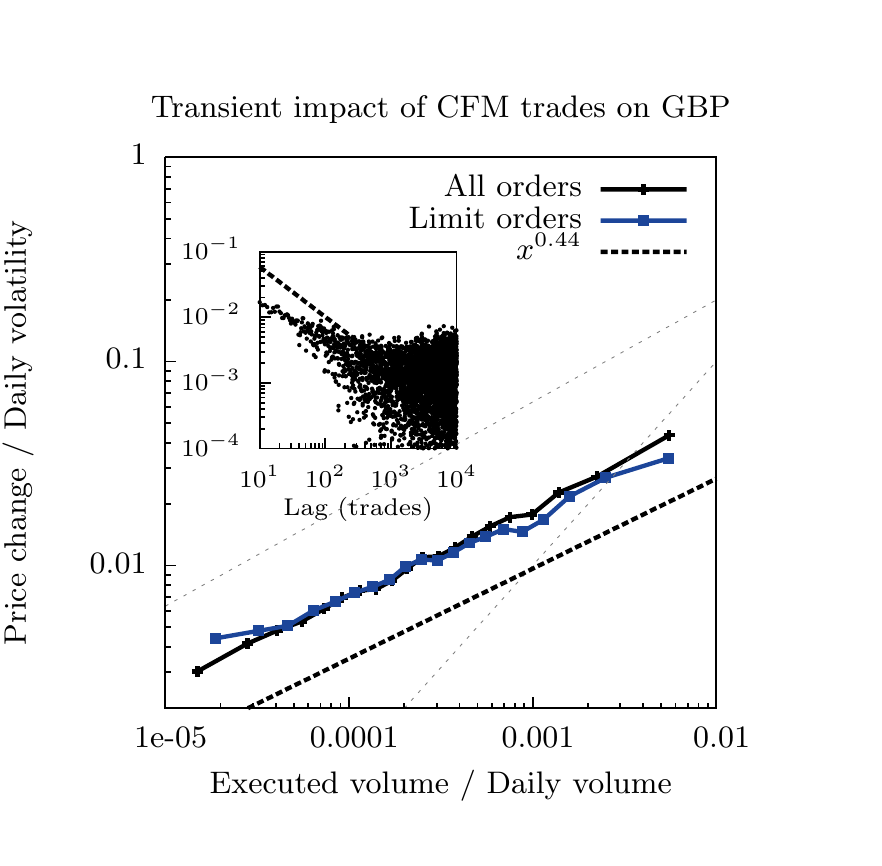}
  \caption{Impact of CFM trades on the GBP futures market, obtained by averaging over $3\times 10^4$ meta-orders executed during the period 2008--2012. The full lines with symbols in the main plot correspond to two styles of execution of the meta-order (either with a mix of limit and market orders, or exclusively with limit orders). The average impact in the two cases appears to be the same. The soft dashed lines plotted for comparison show power-laws with exponents 0.5 and 1. The thick dashed line indicates the result of a power-law fit, with exponent $\delta=0.44$. In the inset we plot the intra-day sign autocorrelation function for the same product averaged over all the trades of year 2012, exhibiting a power-law decay with exponent $\gamma \approx 0.76$.}
  \label{fig:DISCUSImp}
\end{figure}

\begin{figure}[hbt]
  \centering
  \includegraphics{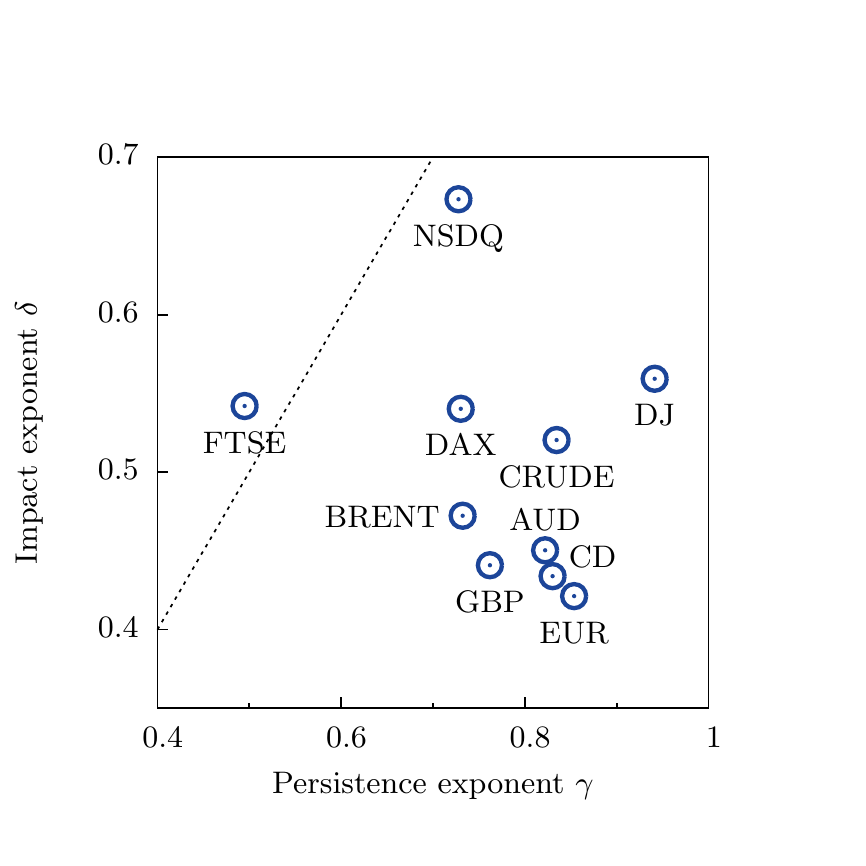}
  \caption{Plot of the impact exponent $\delta$ against the persistence exponent $\gamma$ characterizing the autocorrelation of trade signs for a representative set of futures contracts. The impact exponent $\delta$ has been computed by using a dataset consisting of approximately $10^6$ meta-orders, executed during the period 2008-2012. Buy orders and sell orders lead to very similar results. The exponent $\gamma$ is calculated by using intra-day data for the year 2012, consisting of roughly $10^4$ trades per day per contract (market-wide). Note that there not seem to be any significant correlation between $\delta$ and $\gamma$, in disagreement with the model of \cite{Farmer:2011}, which predicts $\delta=\gamma$ (dashed line).}
  \label{fig:GammaVSDelta}
\end{figure}

This finding is in our opinion truly remarkable, on several counts. First, because it is so universal (across markets and execution
strategies) and stable over time, it is indeed tempting to call it a ``law'' akin to physical laws. There are in our experience 
not that many stable empirical relations in financial markets, and this is one of the most compelling that we have encountered. Second, a square-root dependence is totally counter-intuitive,
at least at first sight, because it is {\it non-additive}. In other words, a square-root law entails that the last $Q/2$ trades have an impact that is only $\sim 40 \%$ of the first $Q/2$! But surely if the last $Q/2$ are executed a year (say) after the first $Q/2$, impact should become additive again. This means that the only possibility for such a strange behaviour to hold is that there must exist some memory in the market that extends over a time scale longer than the typical time needed to execute a meta-order. The second ingredient needed to explain the concavity of the square-root impact is that the last $Q/2$ must experience more resistance than the first $Q/2$. In other words, after having executed the
first half of the meta-order, the liquidity opposing further moves must somehow increase. Still, it is quite a quandary to understand how such non-linear effects can appear, 
even when the bias in the order flow is vanishingly small.

In a previous publication~\cite{Toth:2011}, we proposed a minimal model based on the above two intuitive ingredients (memory and 
liquidity increase), in order to rationalize the universal square-root dependence of the impact. Our argument relied on the
existence of slow ``latent'' order book, i.e.\ orders to buy/sell that are not necessarily placed in the visible order book  
but that only reveal themselves as the transaction price moves closer their limit price. 
We showed, using both analytical arguments and numerical simulations of an artificial market, that 
the liquidity profile is V-shaped, with a minimum around the current price and a linear growth as one moves away from that price. This explains why the resistance to further moves increases with the executed volume, and provides a simple explanation -- borne out by numerical simulations -- for the square-root impact~\cite{Toth:2011}. By the same token, a vanishing expected volume available around the mid-price leads to very small trades having anomalously large impact, as indeed reflected by the singular behaviour of the square-root function near the origin. This has led us to the notion of an inherently critical liquidity in financial markets.
The presence of a liquidity funnel localized around the mid-price is in fact a feature which is expected in any micro-structural model encompassing the notions of ordered prices and market clearing, and emerges even in highly stylized ``reaction-diffusion'' models such as \cite{Bak:1997,Tang:1999}. Hence we expect our predictions concerning price impact to be rather general and robust with respect to the precise specification of the model.

Still, there is a large degree of arbitrariness in the choices that we made to construct a statistically efficient artificial market where prices behave as random walks, 
and some questions have been raised about the generality of our results,\footnote{In particular by our friends and colleagues D. Farmer, J. Gatheral \& F. Lillo.} as well as 
on some more subtle points that were not fully clarified in~\cite{Toth:2011}. In particular, our initial model conferred all the ``intelligence'' to liquidity takers, whereas liquidity 
providers were entirely passive and acting randomly. In real markets, however, we know that statistical efficiency is the result 
of a complex ``tug-of-war'' between liquidity takers -- who create trends through the fragmentation of their trades -- and 
liquidity providers -- who attempt to benefit from the correlated flow of market orders by actively increasing the liquidity opposing the flow (see the discussion in \cite{Weber:2005,Bouchaud:2006}). 

The aim of the present paper is to revisit and extend the model and the arguments of Ref.~\cite{Toth:2011}. We start by providing 
more precise numerical results about the phase transition, observed in~\cite{Toth:2011}, between a super-diffusive (trending) market 
and a sub-diffusive (mean-reverting) market as the parameters of the model are varied. This model is actually extremely interesting in its 
own right, as an example of a random walk in an adaptive environment, for which very few exact mathematical results are available. We then study various aspects of the impact 
of a meta-order, in particular how the execution style affects the shape of the impact and how the impact decays after the last trade of the meta-order. We broadly confirm the conclusions of~\cite{Toth:2011}, 
that such an ``$\varepsilon$-intelligence'' model is indeed sufficient to reproduce a concave impact function, as long as the execution of the meta-order takes place on time scales much shorter 
than the renewal time of the latent order book. We show that different execution styles (i.e. aggressive market orders or passive limit orders) hardly affect the shape of the impact function, 
which demonstrates that the universality of impact concavity is a consequence of the coarse-grained properties of the supply function, but not of the details of the microstructure.   

We then introduce a variation of the original setting of the model, by giving a more symmetric role to market and limit orders for
ensuring price diffusion. Limit orders now adapt to the market order flow and explicitly act as buffers against further price moves. This 
specification allows us to reproduce the long range correlation of the sign of limit orders found in market data, which exactly mirrors the long range
correlation of the sign of market orders. In the original version of the model~\cite{Toth:2011}, these limit order correlations are totally absent and 
the confining role of the order book is purely mechanical. We find that the impact function is again concave in this setting, and looks even closer to empirical data.

We end the paper by providing a general theoretical framework in which all the results discussed so far can be qualitatively understood. 
We sketch some analytical calculations that might allow one to go beyond the numerical simulations and compute explicitly the various properties of the model 
(super- or sub-diffusion properties, temporal dependence of the impact function, etc.). The completion of this program is however left for future studies. 

In section \ref{sec:Model} we introduce the model and discuss its properties in the absence of meta-orders, while in section \ref{sec:Impact} we show that a concave impact function is indeed observed 
for different execution protocols. Section \ref{sec:CorrLimit} presents the generalization of the model in which market and limit orders are allowed to interact, while in section \ref{sec:Theory} we construct a theoretical framework in which all the results discussed so far can be qualitatively understood. Finally, we draw our conclusions in section \ref{sec:Conclusions}.

\section{A dynamical model for ``latent'' liquidity}
\label{sec:Model}

\subsection{The building-blocks of the original model}

The basic assumption underlying the arguments put forward in~\cite{Toth:2011} is the existence of a slowly evolving \emph{latent} order book storing the volume that market participants would be willing to trade at any given price $p$. This latent order book is where the ``true'' liquidity of the market lies, at variance with the real order book where only a very small fraction of this liquidity is revealed, and that evolves on very fast time scales. In particular, market making/high frequency trading contributes heavily to the latter but only very thinly to the former.\footnote{It is actually worth noticing that the ``square-root'' impact law has not been much affected by the development of high-frequency trading; this is yet another strong argument in favor of the latent liquidity models.}

This hypothesis is motivated by market data, which demonstrates that only a very small fraction of the volume daily traded on a market is instantly available in the order book~\cite{Bouchaud:2006}.
 The vast majority of the daily traded volume in fact progressively 
reveals itself as trading proceeds: liquidity is a dynamical process, see~\cite{Weber:2005,Bouchaud:2006,Bouchaud:2008}, and for an early study carrying 
a similar message,~\cite{Sandas:2001}. Clearly, traders tend to hide their intentions as long as they can, as they have no incentive in giving away private information too soon by adding orders to the real order book. In fact, the actual decision to trade at a certain
price $p$ in the future could itself be ``latent'': think for example of a mean-reversion algorithm that would decide to sell if the
price ever went up by a certain quantity. We imagine that the volume in the latent order book materializes in the real order book with a probability that increases sharply when the distance between the traded price and the limit price decreases. In particular, we postulate 
below that the latent order book and the real order book coincide at the best quotes.  

The model for the evolution of the latent order book is inspired from ``zero-intelligence'' models for the real order book~\cite{Farmer:2003,Farmer:2005}, but with some additional features that allow the price to be diffusive~\cite{Toth:2011}. In such setting, the latent order book is modeled as a discrete price grid populated by orders of two species (buy or sell) of variable size. Sell orders sit on the left (\emph{bid}) side of the book, while buy orders populate its right (\emph{ask}) side. Such a book is described by specifying for both sides and for each price level how much volume is instantly available for trading, while its time evolution is dictated by three types of stochastic processes:
\begin{itemize}
\item {\bf Depositions}: An investor who becomes potentially interested in buying or selling shares at a price $p$ places a (virtual) limit order of a unit volume at that price level. We suppose that these limit orders arrive at rate $\lambda$ per unit price, which for simplicity we assume to be uniform along the price line.
\item {\bf Cancellations}: Traders might remove orders which were present in the latent order book. 
We assume any order to have the same probability per unit time $\nu$ to be canceled.
\item{\bf Trades}: A buy (sell) market order might hit the book, resulting in a trade which reduces the volume available on the best ask (bid). Obviously, if the volume on a given price level is completely consumed, a price change is instantly triggered. We assume that market orders follow a Poisson process and denote by $\mu$ the rate at which this type of orders arrive in the market. We neglect here the well known activity clustering in financial markets \cite{Hawkes:recent}, but this effect is, we believe, irrelevant for the present problem. The {\it signs} of these
market orders, on the other hand, have long-range correlations, see below. The statistics of the volume of each market order will turn out to play an important role, as discussed in the next 
subsection. 
\end{itemize}
Notice that the choice $\mu=1$ corresponds to measuring time in units of market orders. An estimate of $\mu$ on stock markets leads to
$\mu = 0.1 - 10\, \text{s}^{-1}$. In the following we will often refer to market order time by using the symbol $t$, as opposed to $\tau$ which will label real time. Without loss of generality, we will also take the tick size $w$ (i.e.\ the spacing of the price grid) to be $10^{-2}$ (in, say, $ \$ $). \\

In the present version of the model, we assume the deposition rate of limit orders to be independent of the side of the book and the cancellation rate to be independent of the sign of the order (but see section \ref{sec:CorrLimit} below). The sign of market orders, on the other hand, is determined by a non-trivial process, such as to generate long-range correlations, in agreement with empirical findings~\cite{Bouchaud:2008}. More precisely, the sign $\epsilon_t$ of market order number $t$ has zero mean (in the absence of a meta-order that would lead to a locally biased flow),  $\langle \epsilon_t \rangle=0$, but is characterized by a power-law decaying autocorrelation function: $\langle \epsilon_t \epsilon_{t^\prime}\rangle = g_{t-t^\prime} \propto |t-t^\prime|^{-\gamma}$ with $\gamma <1$.\footnote{Specifically, we consider the order generation prescription of Ref.~\cite{Lillo:2005}: we use a power-law distribution $p(L) \sim L^{-\gamma-1}$ for the durations $L$ of trends of buy or sell market orders. By taking the sign of each trend to be positive or negative with equal probability we can obtain an autocorrelation function $\langle \epsilon_t \epsilon_{t^\prime}\rangle \sim |t-t^\prime|^{-\gamma}$. While the short-time properties of the model might depend on this choice, its long-range behavior should be independent of the details of the order generation mechanism as long as the asymptotic properties are the same.} This choice is motivated by empirical evidence showing long range correlation in the market order flow, favoring an exponent $\gamma \approx 0.5$ for stock markets and $\gamma \approx 0.8$ for futures markets (see for example Refs.~\cite{Bouchaud:2008,Bouchaud:2004,Lillo:2004} and, for futures markets, Fig.~\ref{fig:GammaVSDelta}).

Another ingredient of the model is the statistics of the volume consumed by each single market order. One possible choice is to assume, as in~\cite{Farmer:2003,Farmer:2005}, that each market order is of unit volume. However, this is unrealistic, since more volume at the best is an incentive
to send larger market orders, in order to accelerate trading without immediately impacting the price. It is more reasonable to posit that the size of market order
$V_{\textsmall{m.o.}}$ is an increasing function of the prevailing volume at the best $V_{\textsmall{best}}$. We proposed in~\cite{Toth:2011} to set 
$V_{\textsmall{m.o.}} = \max{(\lfloor fV_{\textsmall{best}}\rfloor,1)}$, where $\lfloor \dots \rfloor$ means taking the integer part and $f$ is a random variable in $[0,1]$, with a distribution $P(f)$ given by:
\be
P(f) = \zeta (1-f)^{\zeta-1} \; .
\ee
In this way it is possible to tune the aggressivity of market orders through a single parameter $\zeta$ which allows one to interpolate between the case where 
each market order has a unit volume ($\zeta = \infty$) and the case where each market order completely exhausts the volume at the opposite best ($\zeta = 0$).
Intuitively, large values of $\zeta$ (i.e.\ small volumes for each trade) decrease the impact of each trade, 
and therefore the volatility of the market (for a related discussion, see~\cite{Wyart:2008}). 

The cancellation rate defines a time scale $\tau_\nu = \nu^{-1}$ which is of crucial importance for the model, since this is the memory time of the market. For times much larger than $\tau_\nu$, all limit orders have been 
canceled and replaced elsewhere, so that no memory of the initial (latent) order book can remain. Now, as we emphasized above, a concave (non-additive) impact law can only appear if some 
kind of memory is present. Therefore, we will study the dynamics of the system in a regime where times are  small compared to $\tau_\nu$. From a mathematical point of view, rigorous statements about the diffusive
nature of the price, and the non-additive nature of the impact, can only be achieved in the limit where $\nu/\mu \to 0$, i.e.\ in markets where the latent liquidity profile changes on a time
scale very much longer than the inverse trading frequency. Although $\tau_\nu$ is very hard to estimate directly using market data,\footnote{Remember again that $\nu$ is {\it not} the cancel
rate in the real (visible) order book, which is extremely high, $10\, \text{s}^{-1}$ or so, but the cancel rate of trading {\it intentions} in the latent order book, which are much slower.} 
it is reasonable to think that trading decisions only change when the transaction price changes by a few percent, which leads to $\tau_\nu \sim$ a few days in stocks and futures markets. 
Hence, we expect the ratio $\nu/\mu$ to be indeed very small, on the order of $10^{-5}$, in these markets.

\subsection{Super-diffusion vs. sub-diffusion of prices}

We first investigate the statistics of price changes in our artificial model, in particular the variogram $D(t)$ defined as:
\be
D(t) = \langle \left(p(t_0+t) - p(t_0)\right)^2 \rangle,
\ee
where the averaging is either over $t_0$ for a single trajectory (as with real empirical data), or over different trajectories for a given $t_0$ -- but in 
both cases $t_0$ must be chosen $\gg \tau_\nu$ in order to be in the stationary state. A useful quantity is  $\sigma^2_t= D(t)/t$ (the so-called ``signature plot''), 
which can be seen as a measure of the squared volatility on time scale $t$. For a purely diffusive process (e.g.~the Brownian motion), $\sigma^2_t$ is strictly
independent of $t$. A ``sub-diffusive'' process is such that $\sigma^2_t$ is a decreasing function of $t$, signalling mean-reversion, whereas a ``super-diffusive'' 
process is such that $\sigma^2_t$ is an increasing function of $t$, signalling trends. A simple example is provided by the fractional Brownian motion, which is such that $D(t) \propto t^{2H}$, 
where $H$ is the Hurst exponent of the process. 
The usual Brownian case corresponds to $H=1/2$; $H > 1/2$ (resp.\ $H < 1/2$) is tantamount to super- (resp. sub-) diffusion.
It may also happen that the process becomes diffusive at long times, i.e., $\sigma_t^2$ tends to a finite, non-zero value $\sigma^2_{\infty}$ when $t \to \infty$. 

The above dynamical liquidity model contains an ingredient that favors super-diffusion (the long range correlated nature of the order flow), and an 
ingredient that favors sub-diffusion (the long memory time of the order book itself). Let us be more explicit. When the memory time of the order book
$\tau_\nu$ is very short, the autocorrelation of the price changes is dominated by the autocorrelation of the order flow. It is easy to show that for a
power-law autocorrelation with exponent $\gamma$, as defined above, the Hurst exponent of the price change is given by:
\bea
H = \frac12, &\quad& {\mbox{ \text{when}}} \quad \gamma > 1 \\
H = 1 - \frac\gamma2 > \frac12, &\quad& {\mbox{ \text{when}}} \quad \gamma < 1 \, ,
\eea
with logarithmic corrections for $\gamma = 1$. The same result would in fact hold for an arbitrary memory time $\tau_\nu$, 
but when market orders always consume the whole volume available at the best quote (i.e.\ when $\zeta \to 0$). 
Indeed, in this case, each market order  induces a mid-point change. 
For an {\it uncorrelated order flow} (i.e. $\gamma \to \infty$) in a quickly evolving environment ($\tau_\nu \sim \mu^{-1}$) the price process is obviously diffusive. 

However, the same situation of a totally uncorrelated order flow $\gamma \to \infty$, but now with a very slowly evolving order book $\tau_\nu \gg \mu^{-1}$, turns out to be far from trivial, and 
leads to a strongly sub-diffusive dynamics.\footnote{This is actually the reason why the early zero-intelligence models of \cite{Farmer:2003} were unsatisfactory as models of true prices. As can be seen
in Fig. 11 in that paper, the price is indeed strongly sub-diffusive within these models. The need for a correlated order flow to counter-balance this effect was in fact duly noted in the 
conclusion of that paper.} Intuitively, 
this strong mean-reverting behaviour is explained by the following argument: imagine that the price has been drifting upwards for a while. The buy side of the book, 
below the current price, has had little time to refill yet, whereas the sell side of the book is full and creates a barrier resisting further increases. 
Subsequent sell market orders will therefore have a larger impact than buy orders, pushing the price back down. 

When $\zeta \to 0$ and $\gamma \to \infty$, then $H=1/2$ trivially since each market order removes the best quote completely, killing the mean-reversion effect. 
However, in the limit $\zeta \to \infty$ where the executed volume is unity, simulations show that the price motion is actually {\it confined}, i.e.:
\be \label{confined}
\sigma^2_t \approx \sigma^2_\infty - \frac{c}{\sqrt{t}},
\ee
in the double limit $t \to \infty$, $\nu \to 0$ with $\nu t \to 0$, and where $c$ is a constant depending upon the values of $\mu$ and $\lambda$. This result can be intuitively understood after realizing that $\sigma_\infty$ is proportional to the stationary value of the spread: in this regime the price bounces indefinitely  in the region of the bid-ask spread, whereas the volumes of price levels outside that zone grow linearly in time, leading to a trapping effect. \\

When $\zeta$ is finite (neither zero nor infinite), an analogy with diffusion-reaction models, studied in~\cite{Bak:1997,Tang:1999,Maslov:2000}, suggests $H = 1/4$ in the regime $\tau_\nu \gg \mu^{-1}$. Within that framework, buy and sell orders are described as particles of type $A$ and $B$ diffusing along the infinite 
price line. The market clearing condition is then modeled by assuming that buy and sell orders annihilate each other as soon as they meet ($A+B \to \emptyset$), the interface between the $A$-rich region and the $B$-rich region defines the position of the mid-point. This diffusion-annihilation problem has been studied in details, and previous numerical results suggest an exponent $H=1/4$~\cite{Cornell:1995}. More formal arguments building on the results of~\cite{Cardy:1996} can be found in~\cite{Tang:1999,Eliezer:1998}. In our model, we have carefully revisited the limiting case $\tau_\nu \gg \mu^{-1}$, and found  a result compatible with $H = 1/4$ with logarithmic corrections for all finite values of $\zeta$, as predicted in the diffusion-annihilation model~\cite{Cardy:1996}.\footnote{A power-law fit of the results however leads to $H \approx 0.32$, very similar to the numerical results obtained for diffusion-annihilation problem~\cite{Cornell:1995}.} Note that in the order book model considered here, orders do not ``diffuse'' directly, so the mapping to a diffusion-annihilation problem, if correct, does not seem to be trivial. We have not been able to construct a convincing argument that would show that 
the two models are in the same universality class, and the similarities between the two could be misleading.

\subsection{Diffusive prices and market efficiency}

From a financial point of view, both super-diffusion and sub-diffusion lead to arbitrage opportunities, i.e.\ strategies that try to profit from the trends or 
mean-reversion patterns that exist when $H \neq 1/2$. The trading rules that emerge must be such that simple strategies are not profitable, i.e.\ prices are close
to random walks with $H \approx 1/2$, a property often called ``statistical efficiency''.\footnote{Whether or not this random walk behaviour is indicating the 
markets are ``efficient'' in the sense that prices reflect fundamental values is another matter. We believe that while the former property is indeed obeyed, the
mechanisms that lead to statistical efficiency have little to do with the activity of fundamental arbitrageurs. See~\cite{Bouchaud:2008} for an extended discussion
of this point.} Within the original setting of our model, and much in the spirit of~\cite{Bouchaud:2006,Bouchaud:2004}, we have two parameters $\gamma$ and $\zeta$
to play with, that allow us to tune the relative strength of trending effects induced by market orders and the mean-reversion forces induced by limit orders. Note that
since $\gamma$ is empirically found to be smaller than unity, {\it markets must clearly operate in the regime} $\tau_\nu \gg \mu^{-1}$, otherwise a super-diffusive behaviour with 
$H=1-\gamma/2$ would necessarily ensue. 

\begin{figure*}[htb]
  \centering
  \includegraphics{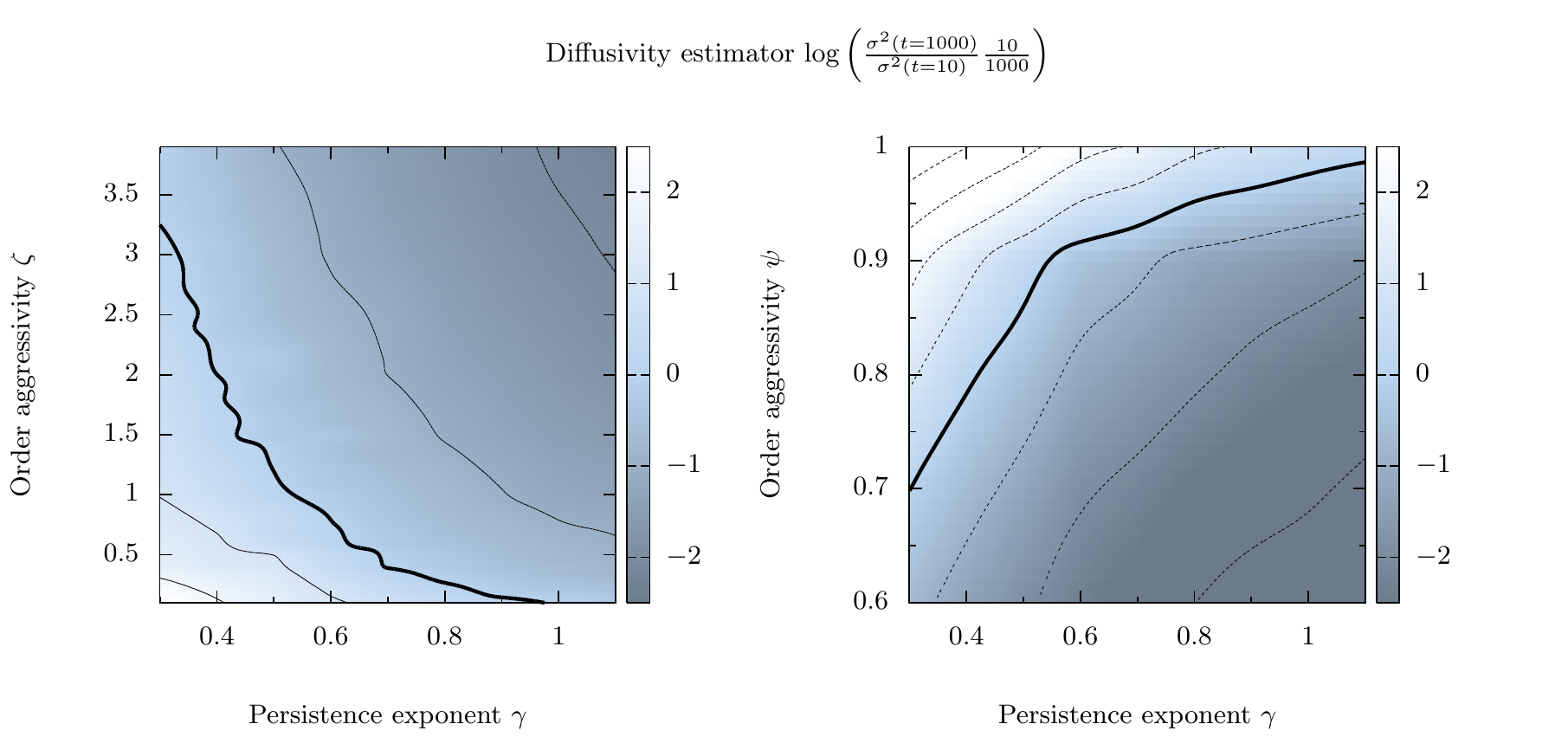}
  \caption{(\emph{Left})~Phase diagram for the model in the regime $\mu=0.1\, \text{s}^{-1}$, $\lambda w = 5 \times 10^{-3} \, \text{s}^{-1}$, $\nu = 10^{-7} \, \text{s}^{-1}$. 
  The diffusive nature of the model is assessed by considering the quantity $S=\log \left( \frac{1}{100}\frac{D(t=10^3)}{D(t=10^1)} \right)$, 
  so that the perfectly diffusive regime diffusion corresponds to $S = 0$ (thick black line). (\emph{Right})~The same quantity is plotted for the modified model in which the order consumption is controlled by the $\psi$ exponent for the same set of parameters. In both cases, for any value of $\gamma$ one can find a critical $\zeta$ beyond which 
  the behavior of the model passes from super-diffusion to (apparent) sub-diffusion.}
  \label{fig:PhaseDiagram}
\end{figure*}

The claim made in~\cite{Toth:2011} is that there exists a critical line, in the $\gamma,\zeta$ plane, separating a phase where the price is super-diffusive 
(below that line) from a phase where it is sub-diffusive (above that line). The artificial market is therefore found to be only ``viable'' (i.e.\ 
efficient) along a certain line $\zeta_c(\gamma)$ in parameter space. Numerically, the location of this critical line is approximately determined
by comparing $\sigma^2_t$ for $t= 10^{1} \mu^{-1}$ and $t=10^{3} \mu^{-1}$~\cite{Toth:2011}. This leads to the result shown in Fig.~\ref{fig:PhaseDiagram}, with new
data.

\begin{figure*}[htb]
  \centering
  \includegraphics{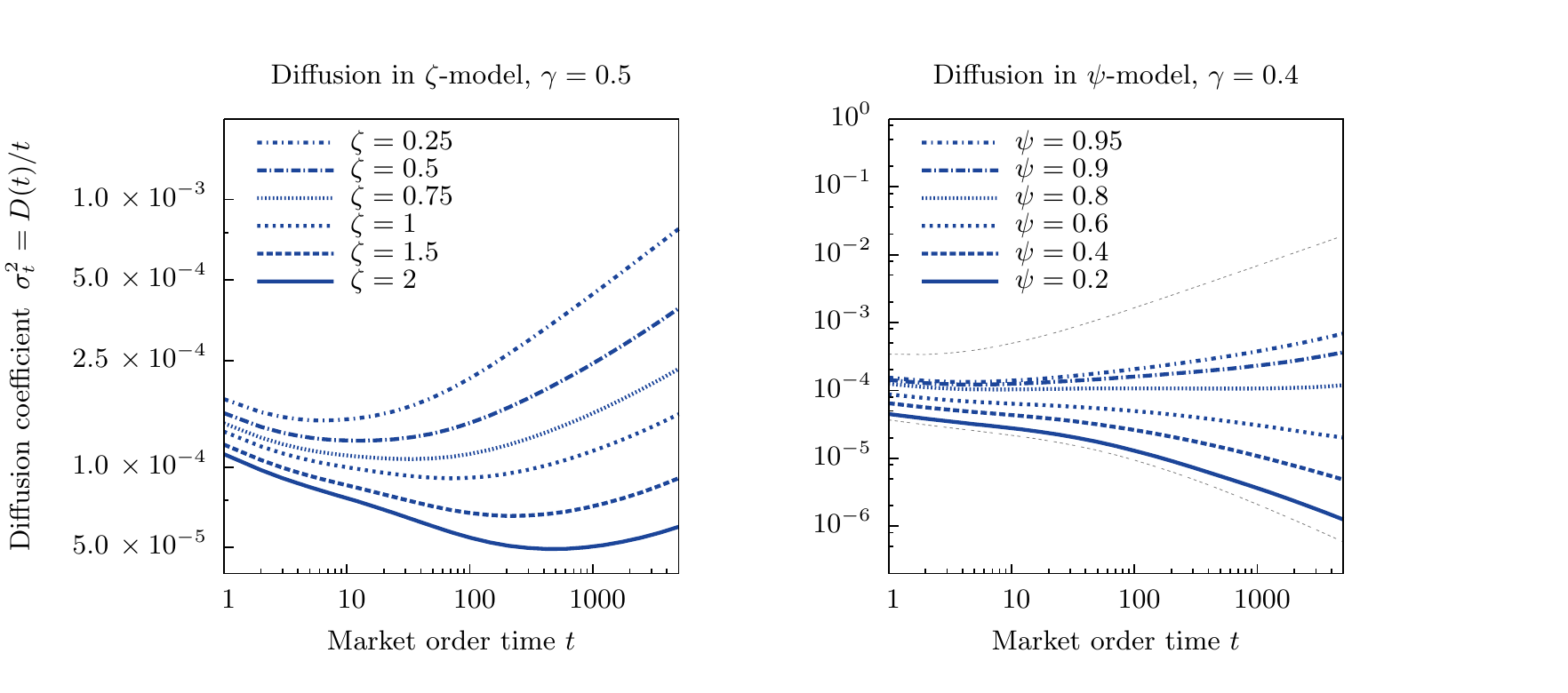}
  \caption{(\emph{Left})~Signature plots $\sigma^2_t$ for the parameter choice $\mu=0.1\, \text{s}^{-1}$, $\lambda w = 5 \times 10^{-3}\, \text{s}^{-1}$, $\nu=10^{-7}\, \text{s}^{-1}$, $\gamma = 0.5$ and different values of $\zeta$. Decreasing values of $\zeta$ lead to more strongly diffusive behavior. Note however that the long-time behaviour is always super-diffusive in this case. (\emph{Right})~Signature plots for the modified model with power law volume consumption, for various values of the $\psi$ exponent. The parameters set is the same one adopted in the left plot, except for the value of $\gamma = 0.4$. The light grey lines correspond to the limiting cases $\psi=0$ and $\psi=1$. Note that in this case, the long time behaviour is sub-diffusive when $\psi \lesssim 0.8$, 
  super-diffusive for larger values of $\psi$ and exactly diffusive for $\psi=\psi_c \approx 0.8$.}
  \label{fig:SignPlot}
\end{figure*}

Mathematically, the claim of~\cite{Toth:2011}, is the following: for any $\gamma < 1$, 
there is a critical value $\zeta_c(\gamma)$, such that the behaviour of the price in the {\it intermediate asymptotics} regime $\mu^{-1} \ll t \ll \tau_\nu$ 
evolves from being super-diffusive ($\sigma^2_t \propto t^{2H}$ with $H > 1/2$) for $\zeta < \zeta_c(\gamma)$ to sub-diffusive ($\sigma^2_t \propto t^{2H}$ with $H<1/2$) 
for $\zeta >\zeta_c(\gamma)$. We have redone extensive numerical simulations of the original model, and our 
conclusion is different. Although there is indeed a value of $\zeta$ for which the price is approximately diffusive for long enough times to
be of practical significance, we find that the super- $\to$ sub-diffusion transition is in fact only a cross-over. Running careful simulations for
long enough times (while still in the intermediate regime $\mu^{-1} \ll t \ll \tau_\nu$), we see that the effect of long-range correlations in the
signs of the trades {\it always dominates} at long times, and lead to an asymptotic Hurst exponent $H= 1 - \gamma/2$ whenever $\gamma < 1$ -- see Fig.~\ref{fig:SignPlot}.

One should nevertheless separate issues of practical interest from purely theoretical questions. From a practical point of view, the approximately diffusive behaviour
of prices observed in~\cite{Toth:2011} over a certain time frame was enough to qualify the model as realistic, and this allowed us to move on to the measure of the impact of meta-orders 
over the same time frame -- as we do in the next paragraph. Still, it is a matter of considerable theoretical interest to define a model where a \emph{super- to sub-diffusion transition} 
truly exists in a mathematical sense. For this purpose, we slightly modify the rule that sets the size of market orders as follows: given a specific sign for a market order,
its volume is set to be a power $\psi$ of the volume on the opposite best $V_{\textsmall{best}}$:
\be
V_{\textsmall{m.o.}} = \max(\lfloor V_{\textsmall{best}}^\psi \rfloor,1) 
\ee
with $\psi \in [0,1]$, so that $V_{\textsmall{best}} \geq V_{\textsmall{m.o.}}$. Clearly, larger values of $\psi$ correspond to more aggressive orders, so that for $\psi=0$ one recovers unit execution, and the price is confined (see Eq.~(\ref{confined}) above). If $\psi=1$, on the other hand, price is trivially super-diffusive when $\gamma < 1$. But when $\psi < 1$, the volume eaten by 
market orders is (asymptotically) a very small fraction of the volume at best, suggesting that the confining effect of the book might end up dominating the
dynamics for small enough $\psi$. This is what we find numerically -- see Figs.~\ref{fig:PhaseDiagram} and~\ref{fig:SignPlot}. More precisely, we now find that the Hurst exponent $H$ of the price process
is a \emph{continuously varying function} of $\gamma$ and $\psi$, monotonically increasing from $H(\gamma,\psi=0)=0$ to $H(\gamma,\psi=1)= \min(1 - \gamma/2,1/2)$.
For all $\gamma < 1$, there is therefore a critical value $\psi_c(\gamma)$ such that market efficiency is strictly recovered for $\mu^{-1} \ll \tau \ll \tau_\nu$.
For $\tau \gg \tau_\nu$ and $\gamma < 1$ we expect, as explained above, super-diffusion to take over in all cases, so our artificial market tuned to be 
efficient on intermediate time scales would still show long term trends. The model is however, obviously, an approximate description of reality, which
neglects many effects that play an important role on longer time scales. One is that the long memory in sign trades is probably cut-off beyond some time scale, 
although this is very difficult to establish empirically. Note also that it is actually the case that weak trends exist in financial markets!

\subsection{The volume distribution at the best} 

\begin{figure*}[htb]
  \centering
  \includegraphics{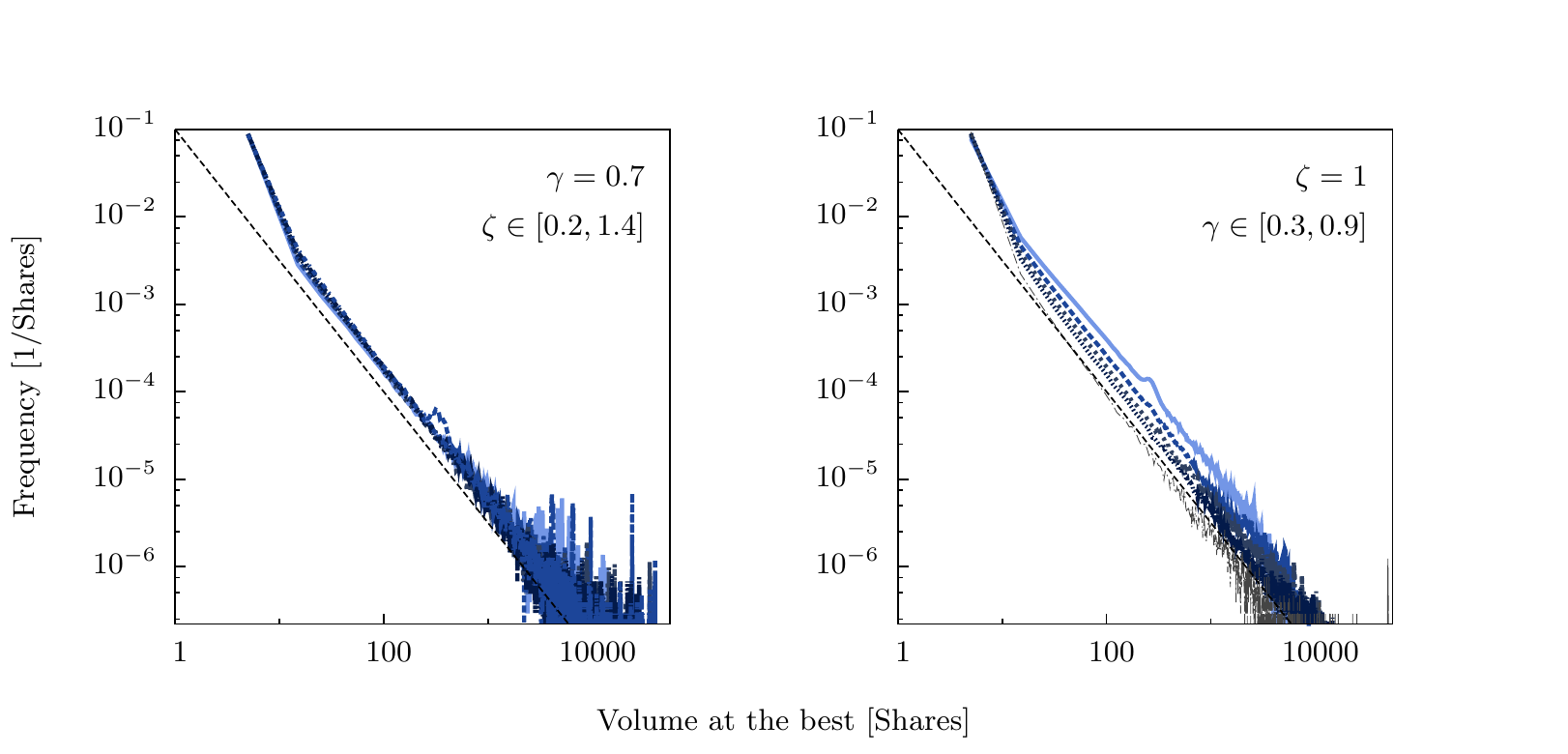}
  \caption{Histogram of the volume distribution for the best bid and ask prices, 
  computed for various values of $\zeta$ (left plot) and of $\gamma$ (right plot). The curves on the left plot are almost exactly superposed, while the ones on the right plot share approximatively the same slope. The grey points on the right plot correspond to the case of independent market orders. 
  The lines plotted for comparison show a power-law decay with exponent $-1.5$. The parameters used to obtain this figures correspond 
  to the ones used to generate Fig.~\ref{fig:PhaseDiagram}, so that the maximum depth is given by $\lambda w /\nu = 5 \times 10^4$.}
  \label{fig:VolumeOnBest}
\end{figure*}

Finally, we have studied the volume distribution at the best bid (or the best ask). 
The results are summarized in Fig.~\ref{fig:VolumeOnBest}, where we plot the empirical distribution that we obtain for various values of $\gamma$ and $\zeta$. 
The volume distribution is very broad, and decays as a power law with a universal exponent (independent of $\gamma$ and $\zeta$) with a value close to $-1.5$. 
One in fact finds a peak in the probability of very large volumes, associated with price levels which have never been hit yet by market orders, 
that have therefore an average volume equal to $\lambda w / \nu$, where $w$ is the tick size.
We can therefore conclude that the broad distribution of volumes is induced by the large depth of the book, 
which forces the price process to visit price bins with volumes ranging from values close to zero (found in the region of the spread) 
to values around the maximum depth (for yet unexplored price regions).

We find that the distribution of queue durations is also broad, which is in our case due to the persistence of the market order flow. Intuitively, a bid queue can survive a long period of time if a long stream of orders hits the ask queue, but as soon as a sequence of orders of the appropriate sign is started, the bid is emptied after a finite number of orders (typically of the order of $\log \lambda/\nu $). However, we have not found a way to derive the apparently universal value~$-3/2$ of the tail exponent for the distribution of the volume at best.

\section{The concave impact of meta-orders}
\label{sec:Impact}

We now revisit the numerical results of~\cite{Toth:2011} for the impact of meta-orders, and study more precisely the dependence of this impact on the style and 
``aggressivity'' of the execution schedule. 

\subsection{Execution with market orders}

We first define more precisely how meta-orders are introduced in the model, on top of the previously defined ``background'' order flow that builds an unbiased,
diffusive price time series. Our choice is to introduce an extra agent (the ``trader'') into the market, which buys (without loss of generality)  $Q$ shares within the time interval 
$[0,T]$, by executing {\it market orders} at a fixed time rate $\mu \phi$ (the case of limit order execution is discussed later). 
The flow of market orders is now biased, with $\langle \epsilon_t\rangle = \varphi$, where $\varphi$ is an increasing function of $\phi$ usually called \emph{participation rate}.
The relation between the participation rate $\varphi$ and the frequency $\phi$ depends in general on the order flow of both the trader and the background, and reduces to $\varphi= \frac{\phi}{1+\phi}$ when on average the trader and the rest of the market submit individual market orders of the same volume. 
After time $T$, the meta-order ends, and the market order flow immediately reverts to its unperturbed state.

We first keep the original setting of T\'oth et al.~\cite{Toth:2011} and work in a region of the plane $\gamma, \zeta$, and for a range of time scales such that the 
price is approximately diffusive. 
(In fact, a concave impact can also be observed even in the super- and sub-diffusive phases.)

We allow the trader to submit orders with a volume extracted by a different distribution with respect to the one used by the rest of the market, 
in order to model different execution styles for the submission of the meta-order. Accordingly, we introduce a quantity $\zeta^\prime$ determining the 
volume consumption of the trader: analogously to the case of $\zeta$, we suppose that the fraction of volume $f^\prime$ consumed by any of the market orders 
submitted by the trader is distributed according to $p(f^\prime) = \zeta^\prime (1-f^\prime)^{\zeta^\prime-1}$.
This implies that for all values of $\zeta^\prime \neq \infty$, whenever the executed quantity $Q$ is fixed, the execution time $T$ fluctuates according to the actual liquidity conditions of the market. Conversely, choosing a submission protocol with fixed $T$ requires $Q$ to fluctuate. Only in the unit execution case $\zeta^\prime=\infty$ it is possible to fix at once $Q$ and $T$.

Because of the bias in the order flow, the average price change $\langle p_T - p_0 \rangle$ between the start and the end of the meta-order is no longer zero. 
The questions we want to ask are: 
\begin{enumerate}
\item Is the dependence of the \emph{initial impact} ${\mathcal I}=\langle p_T - p_0 \rangle$ on $Q$ concave and how does it depend on the frequency $\phi$? 
\item Does the impact depend on the execution style (parameterized here by $\zeta^\prime$)? 
\item What happens to the price at large times after the meta-order is over (i.e., what is the \emph{permanent} part of the impact $\langle p_\infty - p_0 \rangle$)? 
\end{enumerate}

\begin{figure*}[htpb]
  \centering
  \includegraphics{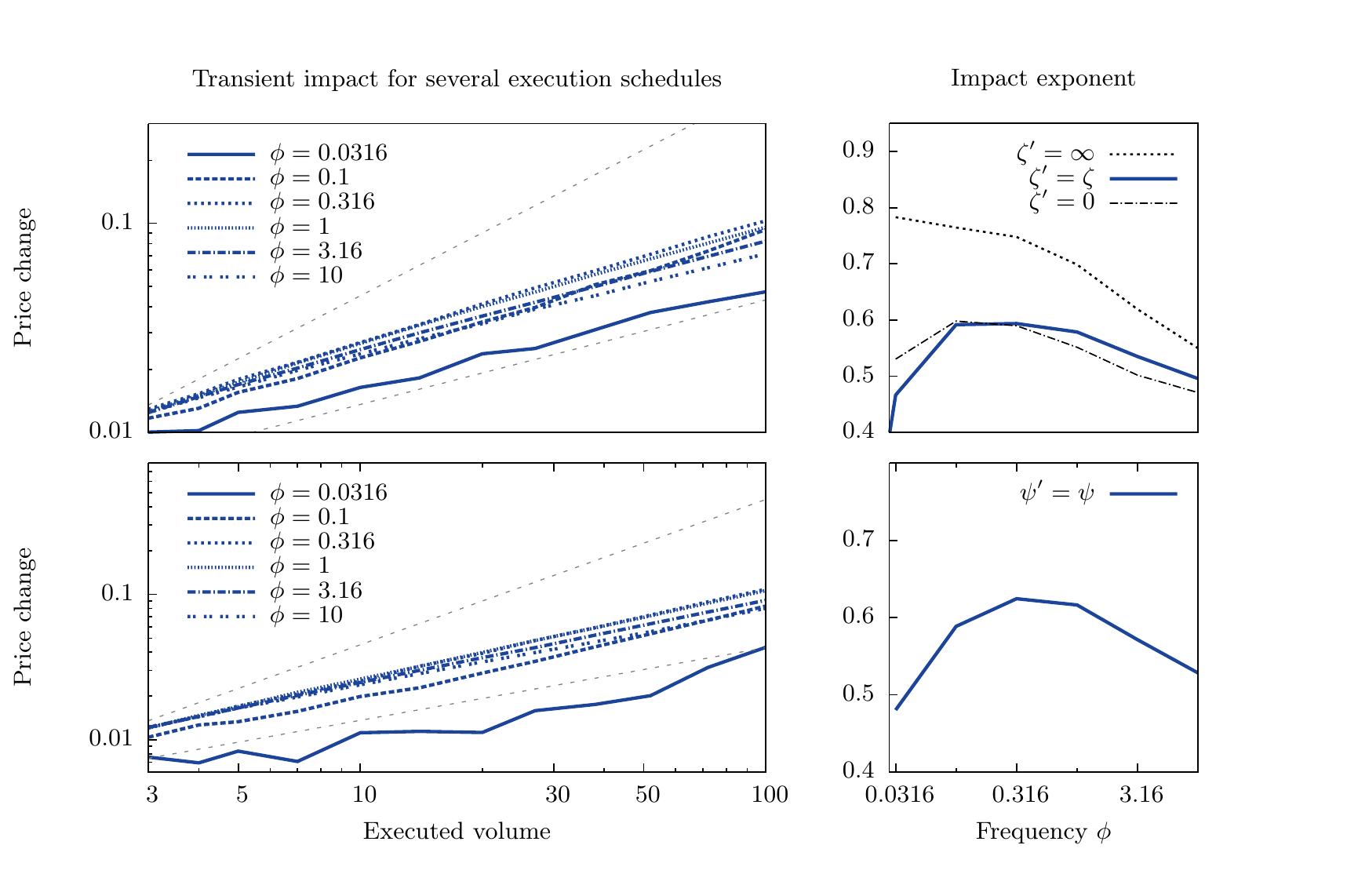}
  \caption{(\emph{Top})~Temporary impact for the execution of a metaorder in the case $\gamma=0.5$, $\zeta^\prime = \zeta=0.95$, for the set of parameters $\mu=0.1 \, \text{s}^{-1}$, $\lambda w = 5 \times 10^{-3} \, \text{s}^{-1}$, $\nu = 10^{-7} \, \text{s}^{-1}$. The right plot shows the fitted exponent for the impact function under this particular execution schedule (solid line), compared with the ones corresponding to different execution protocols (dashed lines). Note that except for unit execution where concavity is weaker, the value of the impact exponent $\delta$ is compatible with empirical data.
(\emph{Bottom})~Temporary impact for the modified model in which the $\psi$ parameter controls the order consumption mechanism. We considered the case $\psi^\prime = \psi = 0.75$ and $\gamma=0.4$ for which the model is approximately diffusive. The other parameters are set to $\mu=0.1 \, \text{s}^{-1}$, $\lambda w = 5 \times 10^{-3} \, \text{s}^{-1}$, $\nu = 10^{-7} \, \text{s}^{-1}$. The right plot shows the fitted impact exponent. The results that we obtain for this model are very close to the ones reported above for the $\varepsilon$-intelligence model. The soft dashed lines in the top and bottom left panel are plotted for reference, and indicate the scalings ${\mathcal I} \propto Q^{1/2}$ and ${\mathcal I} \propto Q$.
}
  \label{fig:TempImpZetaExec}
\end{figure*}

We investigated market impact for several values of the $\zeta^\prime$ parameterizing the volume taken with each market order by the trader. 
We have considered execution volumes in the range $Q=[0,100]$, while the trading frequencies $\phi$ have been studied in the range $[0.01,10]$.
For each value of $Q$ and of $\phi$ that has been considered, we have simulated $3 \times 10^4$ realizations of the submission process and 
computed the average price change during and after each execution. In all the cases under investigation, 
the impact ${\mathcal I}_T$ was found to be a concave function of the volume $Q$:
\be
{\mathcal I}_T \propto Q^\delta \qquad \delta < 1,
\ee
confirming the  results reported in~\cite{Toth:2011}.
For example, we plot in Fig.~\ref{fig:TempImpZetaExec} the results obtained for the case $\gamma = 0.5$ and for $\zeta^\prime=\zeta=0.95$. The dependence of the impact exponent $\delta$ on $\varphi$ is shown in the right panel of Fig.~\ref{fig:TempImpZetaExec}, for different values of $\zeta^\prime$.
Note that the impact is significantly less concave in the case of unit
order execution $\zeta^\prime = \infty$ (see right panel of Fig.~\ref{fig:TempImpZetaExec}), whereas $\delta$ is found to be compatible with empirical data, i.e.\ $\delta \approx 0.5 - 0.6$ 
(\cite{Almgren:2005,Toth:2011,Kyle:2012}, and refs.\ therein) for finite values of $\zeta^\prime$, including the case $\zeta^\prime=0$, corresponding to ``greedy'' execution. We also notice that for low participation ratio the impact exponent drops significantly. This is due to a conditioning effect which favors slow execution of buy meta-orders for negative price trajectories and a fast execution in the case of positive price trajectory, causing a bias effect at low $\phi$, as better elucidated in section~\ref{sec:LimitExec}.

In order to check the robustness of the results with respect to the specification of the model, we have simulated the execution of a meta-order also for the $\psi$-model described above, in which the volume consumed by market orders is a power of the volume available at the best. We have chosen the same submission schedule as for the $\varepsilon$-intelligence market, and tested the same range of executed volumes and participation ratios. Even in this case, we find quantitatively very similar results with respect to the previous case, as reported in the bottom panel of Fig.~\ref{fig:TempImpZetaExec}.

\begin{figure}[htbp]
   \centering
   \includegraphics{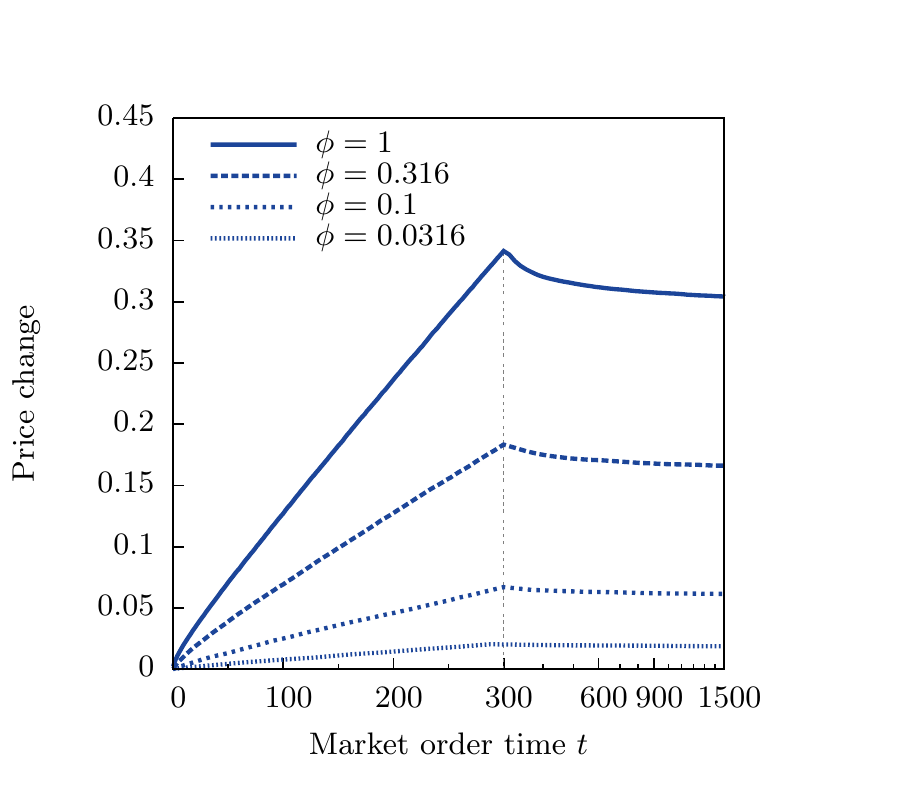}  
   \caption{Dynamics of price impact during the execution of a trade of fixed duration $T=300\,\mu^{-1}$ and stochastic executed quantity $Q$ for different participation rates. We have used $\gamma = 0.5$, $\zeta = \zeta^\prime = 0.95$, $\mu=0.1 \, \text{s}^{-1} $, $\lambda \, w = 5 \times 10^{-3} \, \text{s}^{-1} $, $\nu = 10^{-4} \, \text{s}^{-1} $, while the simulation consisted of $3\times 10^4$ realizations of the submission process. Even though the impact is concave at small times, it finally crosses over to a linear (in time) regime. The relaxation part is represented in semi-logarithmic scale.}
\label{fig:RelaxationEpsSubmission}
\end{figure}

The relaxation of impact after the end of a meta-order is a particularly important topic, which has attracted considerable attention recently. Farmer et al.~\cite{Farmer:2011}
argue that a `fair price' mechanism should by at play, such that the impact of a meta-order reverts at long times to a value precisely equal to the average 
price at which the meta-order was executed (see also \cite{Donier:2012}). This seems to be confirmed by the empirical data analysed in~\cite{Bershova:2013,Waelbroeck:2013}; however, such an analysis is quite tricky, as it involves some degree of arbitrariness in the choice of the timescale for the relaxation of price after the end of the meta-order. Indeed,
we have not been able to confirm this result on CFM's proprietary trades. Even within our synthetic market framework, the long time behaviour of the impact is quite
noisy. Our data suggests that the impact decays to a finite value, which seems to be higher than the `fair price' benchmark, although we cannot exclude a slow decay
to a smaller value. More specifically, we find that permanent and transient component of the impact obey two different scalings: while the transient component of the impact is described by a concave law, its permanent component is linear, and hence dominates the total impact for long enough trades. This behavior can be understood on the basis of the arguments that will be presented in section \ref{sec:Theory}, where we show how the linear component of the impact is initially hidden by a concave transient effect due to the partial adaptation of the order book to the modified order flux.
Overall, we confirm 
again the results presented in~\cite{Toth:2011}, see their Fig.~5 (right). We do also confirm that the initial part of the decay, just after the meta-order is completed, is
very steep, of the type: ${\mathcal I}_{T+t} - {\mathcal I}_T \propto -t^\theta$, with $\theta < 1$. Fig.~\ref{fig:RelaxationEpsSubmission} shows four typical decay curves for market impact for different participation rates.

\subsection{Plasticity of the order book}
\label{sec:BookPlast}
\begin{figure}[htbp]
  \centering
  \includegraphics{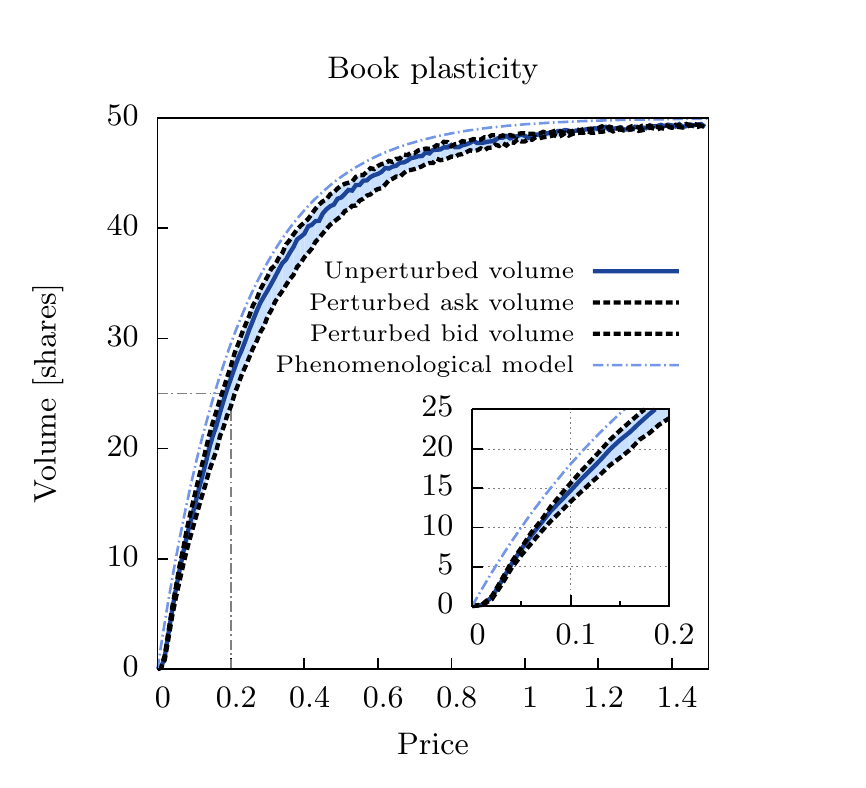}
  \caption{Average shape of the book at equilibrium before (thick solid line) and after (dashed lines) a long buy meta-order executed at a frequency 
  $\phi = 0.0316$ with $\zeta^\prime = \infty$. We considered the same parameters as in Fig.~\ref{fig:TempImpZetaExec}, except for a larger value of $\nu = 10^{-3}\mu$. 
  The soft dashed line indicates the prediction of Eq.~(\ref{eq:MeanFieldEqnBook}) for the unperturbed state of the book. 
  The plot shows that after a long buy meta-order the ask levels are on average more populated than the bid ones. 
  The presence of an offset between bid and ask volume in the perturbed case also evidences that the bulk of the book can store information about the past order flow.}
  \label{fig:AvgBook}
\end{figure}

The shape of the latent order book plays an important role in determining the properties of the model, namely the diffusion behaviour 
and the price impact function discussed above. This will be substantiated more precisely in the last section of this paper (see Eqs.~(\ref{eq:ImpactMarkov}) 
and (\ref{eq:ACMarkov})). The stationary shape of the latent order book when no meta-order is present is represented in Fig.~\ref{fig:AvgBook}, for a choice 
of parameters such that the price dynamics is approximately diffusive ($\gamma=0.5$ and $\zeta=0.95$, while $\mu=0.1 \, \text{s}^{-1}$, $\lambda w = 5 \times 10^{-3} \, \text{s}^{-1}$, $\nu = 10^{-4} \, \text{s}^{-1}$). More generally, we always find that the average book volume 
is an increasing function of the price level $p-p_0$ (where $p_0$ is the current price). The book profile $\rho(p)$ increases from $\rho(p_0)=0$ 
at the mid-price, to the asymptotic value $\rho(\pm \infty) = \lambda/\nu$. The size of the ``liquidity hole'' around $p_0$ is determined by the 
price scale $p^\star \propto \sqrt{\frac{\mu^3}{\lambda^2 \nu}}$. Therefore, the small cancellation limit corresponds to the limit of large latent volume 
$\lambda/\nu \to \infty$ and large liquidity hole $p^\star \to \infty$.

As shown in~\cite{Bouchaud:2002,Toth:2011}, the dynamics of the average shape of the order book can be approximately described, in the diffusive regime, by the equation
\begin{equation}
  \label{eq:MeanFieldEqnBook}
  \frac{\partial\langle\rho(p,t)\rangle}{\partial t} =
 \frac{D}{2} \frac{\partial^2 \langle \rho(p,t) \rangle}{\partial p^2} - \nu \langle \rho(p,t) \rangle  + \lambda \; ,
\end{equation}
where $D$ is the price diffusion constant (or volatility squared). This implies that in the stationary state one has:
\begin{equation}
  \label{eq:MeanFieldShapeBook}
  \langle \rho_\infty(p) \rangle = \frac{\lambda }{\nu} \left( 1- e^{-p/p^\star} \right) \; ,
\end{equation}
where $p^\star = \sqrt{\frac{D}{2\nu}}$. This prediction is compared in Fig.~\ref{fig:AvgBook} against simulated data, with no free fitting parameter.

It is interesting to study how the shape of the order book is progressively deformed by the order flow, and how this 
determines the impact of further trades.  We therefore also show in Fig.~\ref{fig:AvgBook} the asymptotic shape of the latent book 
after a long buy meta-order is executed. One can clearly see how the bid side and the ask side of the book become {\it asymmetric}, while the bid-ask spread remains 
substantially unchanged (inset of Fig.~\ref{fig:AvgBook}). The volume on the ask (sell) side of the book increases, leading to a smaller expected price impact for trades 
in the same direction, and therefore a concave impact. Conversely, the volume on the bid (buy) side of the book decreases, 
which implies that the impact of a sell order coming at the end of a buy meta-order
will be substantially larger than both the impact of an additional buy order and the average impact of buy/sell orders in equilibrium. Hence, impact is expected to relax 
when a buy meta-order stops, because a subsequent balanced order flow will have an average negative impact on the price.  

\subsection{Execution with limit orders}
\label{sec:LimitExec}
The execution of a large meta-order in a real market usually involves a (sometimes large) fraction of limit orders. Interestingly, empirical data indicate that even in this case the impact function is a concave function of the volume, quite comparable to the impact of market order execution (see Fig.~\ref{fig:DISCUSImp}).
This requires any reliable model of market impact to predict concave impact function regardless of the execution protocol (via market, limit orders, or both). This is 
again an indication that market impact is a ``coarse-grained'' effect that depends on the true liquidity and not on market microstructure. This is why we want to
ascertain that the same is true within our numerical model and study the impact of a buy meta-order executed through \emph{limit orders} only.

In fact, modelling limit order execution is more complex than describing market order execution, as the former involves several possible choices: at which price level should the orders be submitted? How should the average lifetime of the orders sitting in the book be fixed? What is the volume deposed with each submission? Our choice is to consider for simplicity a stylized execution strategy for limit orders, mimicking as closely as possible the $\zeta$-execution strategy described above for the case of market order execution. We expect our results not to be strongly sensitive to the precise specifications of the execution protocol.
Accordingly, we have introduced  on top of the unperturbed flow of the $\varepsilon$-intelligence model an extra agent submitting limit orders at the best bid for a volume equal to
\be
V_{\textsmall{l.o.}} = \max(\lfloor f V_{\textsmall{best}} \rfloor , 1)  \; ,
\ee
with a constant, deterministic fraction $f$ of the volume of the best bid.
As in the market order case, we studied depositions occurring at a rate $\mu \phi$. The execution is then interrupted as soon as the cumulated volume executed exceeded a target volume $Q$. We have considered the same background as in the market order submission case (i.e., a market approximately diffusive with $\gamma=0.5$ and $\zeta=0.95$) and averaged our results over $3 \times 10^3$ realizations of the submission process. We have also assumed that orders are never canceled from the \emph{latent} order book, implying that, although the orders might disappear from the real order book, they are reinserted as soon as the book moves close to their original price level.

\begin{figure}[htb]
  \centering
  \includegraphics{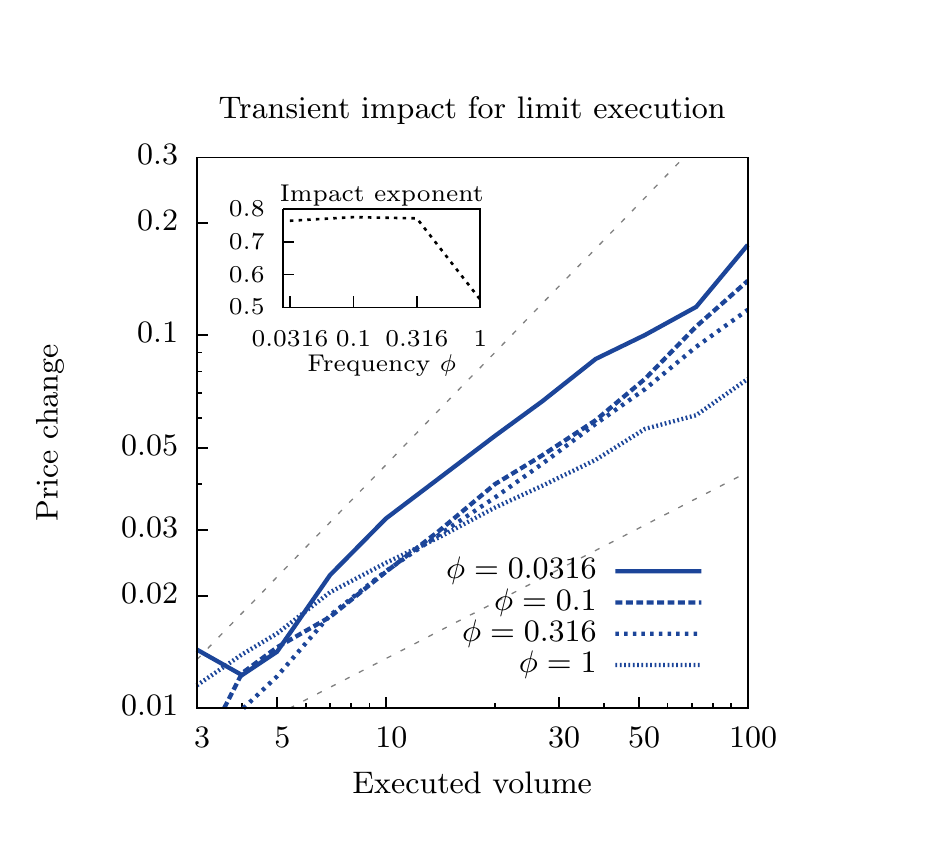}
  \caption{Temporary impact induced by a meta-order executed through the deposition of limit orders. We have considered submissions of volumes equal to a fraction $f=1/2$ of the best, and fixed the other parameters to $\mu=0.1 \, \text{s}^{-1}$, $\lambda w = 5 \times 10^{-3} \, \text{s}^{-1}$, $\nu = 10^{-4} \, \text{s}^{-1}$. As in Fig.~\ref{fig:TempImpZetaExec}, soft dashed lines indicate the reference scalings ${\mathcal I} \propto Q^{1/2}$ and ${\mathcal I} \propto Q$.}
  \label{fig:LimitExecTransImp}
\end{figure}

The remarkable result is that even with limit-order execution we still measure concave impact curves. For intermediate participation rates ($\phi$ from 0.1 to 1), we also found price changes similar to  the ones measured for a market order execution (see Fig.~\ref{fig:LimitExecTransImp})\footnote{
At small execution frequencies ($\phi < 0.1$), the impact is biased by a conditioning effect which has a different sign in the market order execution case and in the limit order one. Specifically, fixing the executed volume $Q$ for a buy meta-order executed through market orders favors negative price trajectories, as a trader with $\zeta^\prime\neq \infty$ tends to wait for the price to uptrend in order to execute large volumes at the ask. The inverse effect is measured for the (buy) limit order execution case, as a negative price swing is required in order to clear orders sitting at the bid. This favors positive price trajectories, thus biasing the price change up.
}. This result indicates that within the present framework market impact should be regarded as a property stemming from the mechanism with which the market clears volume imbalances, rather than a feature depending upon the details about how such imbalances are created. In this respect, the concave shape of market impact should be regarded as a universal feature reflecting the regime of critical liquidity provision in which our synthetic market operates.

\section{An alternative model:  ``stimulated liquidity refill''}
\label{sec:CorrLimit}

The results presented above pertain to the original model of T\'oth et al.~\cite{Toth:2011}, and broadly confirm the main finding of that paper, i.e.\ that the impact of
meta-orders is indeed concave, and in quantitative agreement with empirical data. However, this model has been criticized on the basis that price efficiency is ensured by 
tuning the $\zeta$ parameter that governs the statistics of market orders only, while assuming limit orders to be essentially random and passive. Surely one expects that price efficiency 
actually results from a subtle ``tit-for-tat'' balance between the flow of market orders and the corresponding counterflow of limit orders -- see e.g.~\cite{Weber:2005,Bouchaud:2008,Biais:1995,Eisler:2012} for empirical evidence. Although the aim of~\cite{Toth:2011} was to show that one could build a simple latent order book model that 
is consistent both with price diffusion and with a concave impact, we agree that moving closer to reality is indeed necessary to make the story more compelling.

To that effect, we now present a model where price efficiency is maintained through a ``stimulated liquidity refill'' mechanism, whereby market orders attract a liquidity
counterflow. More precisely, we posit that after a market order of sign $\epsilon$, the probability for the next limit order to be on the ask ($+$) or on the bid ($-$) side of the 
order book is biased as: 
\begin{equation}
  \label{eq:AsymmetryLimit}
  P_\pm( \epsilon) = \frac{1\pm\alpha \epsilon}{2} \; ,
\end{equation}
where $\alpha \in [0,1]$ is a new parameter describing the limit order flow reaction to market orders.
The statistics of the market order flow is still captured by the two parameters
defined above: $0 < \gamma < 1$ describes the long-range correlation of the sign of market orders, whereas $\zeta$ describes the aggressivity of market orders (i.e.\ the fraction of the
opposite volume against which they execute). 

The model studied in the previous sections corresponds to $\alpha=0$, i.e.\ a complete decoupling between market order flow and liquidity provision. For $\alpha > 0$, on the other hand, more volume is on average 
placed on the ask side of the book after a buy market order, and vice-versa for sell market orders. This prescription for limit order deposition reproduces the empirically known correlation between the signs of market orders and limit orders~\cite{Biais:1995}, and the long range correlation of the limit order flow~\cite{Lillo:2004,Eisler:2012}. Notice that even though limit orders are described by a short memory process, the induced correlation between limit orders is effectively long range due to their interaction with market orders.

\begin{figure}[h]
  \centering
  \includegraphics{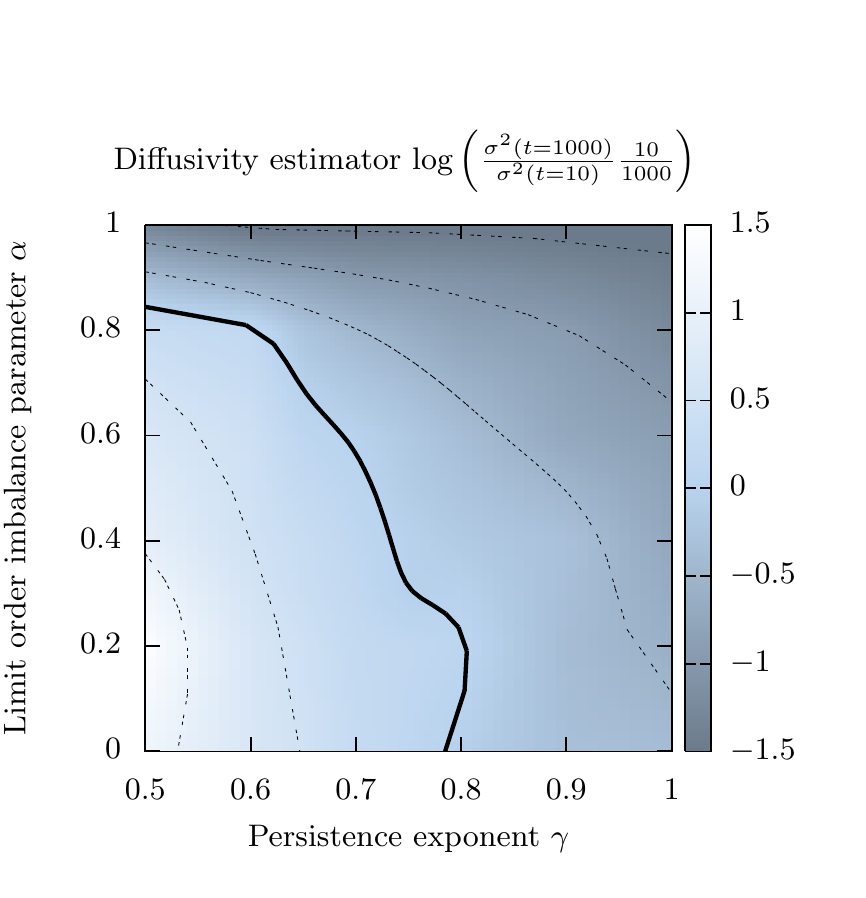}
  \caption{Phase diagram for the model with correlated limit orders in the plane $\gamma, \alpha$. We used the parameters $\zeta = 0.4$, $\mu=0.1 \, \text{s}^{-1}$, $\lambda w = 5 \times 10^{-3}\, \text{s}^{-1} $, $\nu = 10^{-4}\, \text{s}^{-1}$. The diffusion behaviour is estimated as in Fig.~\ref{fig:PhaseDiagram}. We again find, within this setting, a crossover line that separates sub- and super-diffusion regimes.}
  \label{fig:CorrPhaseDiagram}
\end{figure}

Now, choosing a value of $\zeta$ such that for $\alpha=0$ the market is super-diffusive, one can study how increasing values of $\alpha$ progressively decreases the positive autocorrelations induced by the market orders, and eventually leads to an approximately diffusive price for a ``critical'' value of $\alpha$. This is summarized in Fig.~\ref{fig:CorrPhaseDiagram}, where a phase diagram analogous to the one of Fig.~\ref{fig:PhaseDiagram} is shown in the plane $\gamma, \alpha$ for the specific case $\zeta=0.4$, with a crossover value $\alpha_c(\gamma)$ for which the market is approximately efficient.

\begin{figure}[htb]
  \centering
  \includegraphics{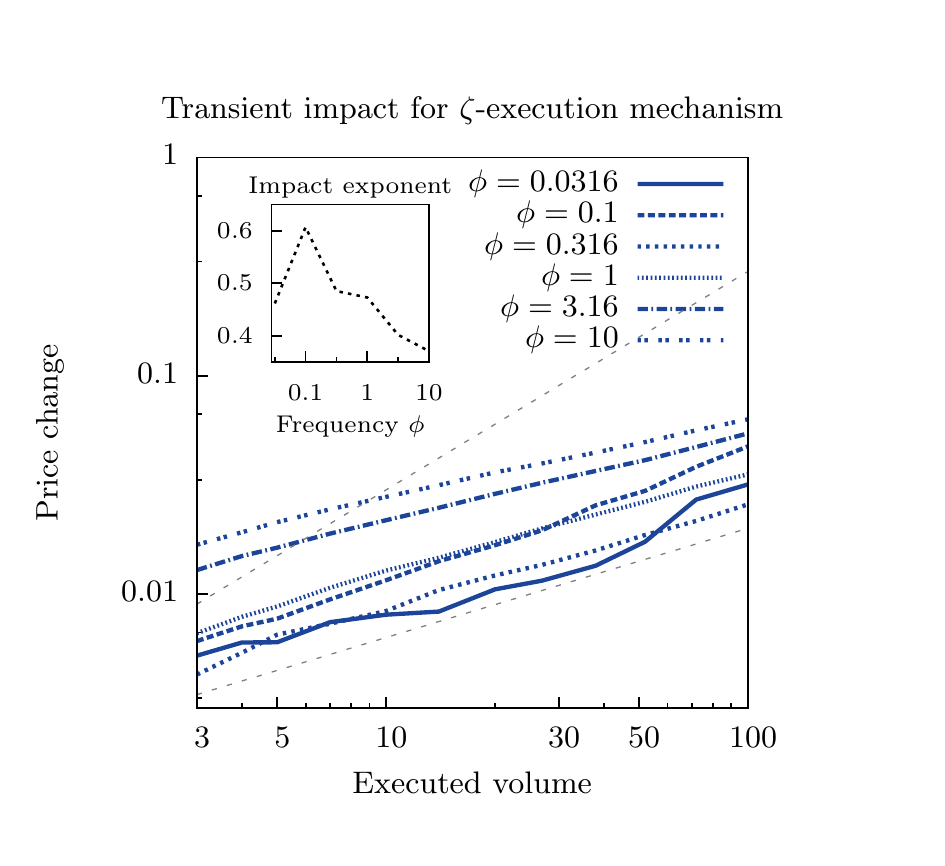}
  \caption{Temporary impact for the execution of a meta-order in a model with correlated limit orders. We have considered $\alpha = 0.85$, $\gamma=0.5$ and $\zeta^\prime = \zeta = 0.4$, corresponding to an approximately diffusive price dynamics. The choice of the other parameters is $\mu=0.1 \; \text{s}^{-1}$, $\lambda w = 5 \times 10^{-3} \, \text{s}^{-1}$, $\nu = 10^{-4} \, \text{s}^{-1}$. As in Fig.~\ref{fig:TempImpZetaExec}, we use soft dashed lines in order to display the reference scalings ${\mathcal I} \propto Q^{1/2}$ and ${\mathcal I} \propto Q$. Impact is strongly concave, with a weak dependence of the impact exponent on the participation rate. The inset shows the dependence of the impact exponent $\delta$ upon the participation rate $\phi$.}
  \label{fig:CorrTempImpZetaExec}
\end{figure}

Finally, we have simulated the execution of meta-orders within the present specification of the model and calculated the corresponding price impact. We again obtain 
a strongly concave shape of the impact as a function of the size of the meta-order, ${\mathcal I}_T \propto Q^\delta$, as shown in the inset of Fig.~\ref{fig:CorrTempImpZetaExec}. Note that
the value of the exponent $\delta \approx 0.4 - 0.5$ only depends weakly on the participation rate.  

The conclusion of this section is that the mechanism that we propose to explain the concavity of the market impact curves is indeed robust to the precise specification of the 
``microscopic'' model of order flow. The advantage of this last version of the model over the original one introduced in~\cite{Toth:2011} is that 
liquidity provision is much more realistic. In particular, the long-range correlation in the sign of limit orders~\cite{Lillo:2004,Eisler:2012} is now correctly reproduced. Interestingly, we find that 
impact is even {\it more concave} than in the original model, with an exponent $\delta \approx 0.4 - 0.5$ for liquidity refill, compared to the value $\delta \approx 0.5 - 0.6$ found above (see Fig.~\ref{fig:TempImpZetaExec}). This is due to the fact that the stimulated liquidity refill mechanism induces a faster volume response to market order trends than in the original model. In fact, the results in the next section will allow us to relate the speed of the response to the exponent $\delta$: we will show that the more prompt is the response, the more concave is the behavior of the impact function.

\section{A general framework for Markovian order books}
\label{sec:Theory}

We now outline a general theoretical framework in which the results presented in the above sections can be qualitatively justified. In order to do this, we construct a faithful representation of our synthetic market allowing us to relate the main properties of the price process to the ones of the order book. In particular we will be able to relate the dynamic properties of the price (e.g., market impact and diffusivity) to the dynamic properties of the first levels of the order book. We will show that the only hypothesis needed to do this is that the evolution of the order book is Markovian. 
Such an hypothesis is trivially fulfilled by our synthetic market (the state of the order book after an event only depends on its shape before the event). Notice that, even though the order book itself stores no memory about its past, the long memory of the sign process $\epsilon_t$ induces indirectly long-ranged correlations on the state of the book.

For the sake of clarity, we will first present the results predicted by the propagator model~\cite{Bouchaud:2004}, which is the simplest setting where one can address the issues of market efficiency and anomalous impact. In fact, even though the predictions of the propagator model and of the above $\varepsilon$-intelligence model are quite different, the structure of the two models are in fact closely related. The analogy between the equations governing price diffusion and market impact for the two family of models will turn out to be useful in understanding some general properties of Markovian order 
book models.

\subsection{Linear models of price impact}
The propagator model assumes that the relation between sign of trades and price changes can be described by a linear relation of the form ~\cite{Bouchaud:2004,Bouchaud:2008} 
\begin{equation}
  \label{eq:PropModel}
  p_t = p_0 + \sum_{t^\prime = 0}^{t-1} G_{t-t^\prime} \epsilon_{t^\prime} \; ,
\end{equation}
where $p_t$ represents the mid-price after trade $t$ and $\epsilon_t$ is a stochastic term denoting the sign of the market order number $t$. The \emph{propagator} $G_{t - t^\prime}$ describes how a trade executed at time $t^\prime$ influences the price at a subsequent time $t$ (i.e., it is assumed that $G_{t - t^\prime}=0$ for $t<t^\prime$). This implies that the properties of the price $p_t$ are completely specified by the properties of the market order flux $\epsilon_t$.
In order to describe the scenario of our synthetic market, we will take
\begin{eqnarray}
  \label{eq:MarkOrderBias}
    \langle \epsilon_t \rangle &=& \varphi \\
\label{eq:MarkOrderCorr}
\langle \epsilon_t\epsilon_{t^\prime} \rangle - \langle \epsilon_t \rangle \langle \epsilon_{t^\prime} \rangle &=& (1-\varphi)^2 g_{t-t^\prime} \; . 
\end{eqnarray}
This allows us to model the long range correlation of market orders through the term $g_{t-t^\prime} \sim |t-t^\prime|^{-\gamma}$ together with the execution of a meta-order with a participation rate $\varphi$. The mean and variance can be easily calculated, and result in:
\begin{widetext}
\begin{eqnarray}
  \label{eq:ImpactPropModel}
  \langle p_t - p_0 \rangle &=& \varphi\sum_{t^\prime = 0}^{t-1} G_{t-t^\prime} \\
  \label{eq:DiffPropModel}
  \langle (p_t - p_0)^2 \rangle - \langle (p_t - p_0) \rangle^2 &=& (1-\varphi)^2\sum_{t^\prime , t^{\prime\prime} = 0}^{t-1} G_{t-t^\prime}G_{t-t^{\prime\prime}}g_{t^\prime -t^{\prime\prime}} \; .
\end{eqnarray}
\end{widetext}
Eq.~(\ref{eq:ImpactPropModel}) expresses the impact of a meta-order executed at the participation rate $\varphi$, while Eq.~(\ref{eq:DiffPropModel}) determines the diffusion properties of the model. In particular the process is diffusive if and only if the sum in Eq.~(\ref{eq:DiffPropModel}) is linear in $t$. This condition fixes the large time behavior of the propagator, relating its shape to the correlation of the order flow: as shown in~\cite{Bouchaud:2004} the propagator model is diffusive if and only if $G_t \sim t^{-\beta}$ with $\beta=\frac{1-\gamma}{2}$ (see also the discussion in \cite{Bouchaud:2008}). 

It will also be convenient to decompose the market impact as
\begin{equation}
  \label{eq:PermTempImp}
  \langle p_t - p_0 \rangle = \varphi \,  G_\infty\, t + \varphi \sum_{t^\prime = 0}^{t-1} (G_{t-t^\prime} - G_{\infty}) \; ,
\end{equation}
in order to distinguish the linear contribution to the impact from the decaying one, which we will call \emph{transient}. Two scenarios for the market impact are compatible with this framework:
\begin{itemize}
\item {\bf The transient impact term is integrable}. The second term then converges to a finite constant and the total impact is dominated at large times by the linear term of equation (\ref{eq:PermTempImp}).
\item {\bf The transient impact term is not integrable}. The total impact at large times is a sum of linear and transient contributions. In particular if the term $G_\infty$ is zero, only the transient component survives.
\end{itemize}
Since the diffusivity of prices requires $G_t \sim t^{-\beta}$, one has $G_\infty = 0$. It results that at large times $\langle p_t - p_0 \rangle \sim \varphi \, t^{\frac{1+\gamma}{2}}$. Inserting $Q = \varphi t$, one then finds for the impact: $\mathcal I \sim Q^{\frac{1+\gamma}{2}} \varphi^{\frac{1-\gamma}{2}}$. Realistic values of $\gamma$ (e.g.~$\gamma\approx 0.5$) then imply that for this model the impact exponent is $\delta\approx 0.75$, with a weak $\varphi^{0.25}$ dependence of the impact on the participation rate, but at odds with the empirical results presented in~\cite{Almgren:2005,Moro:2009,Toth:2011} where $\delta$ is closer to $0.5$. 

In spite of its simplicity, this exercise provides an interesting insight about the origin of anomalous impact: impact is anomalous when the market \emph{slowly adapts} to the presence of a bias in the order flow. Notice in fact that the minimal ingredients in order to have anomalous impact are (i) the fact that the bias $\varphi$ is reduced and eventually absorbed ($\langle \Delta p_t \rangle = \varphi \, G_{t-1} \to 0$ for $t$ large) and (ii) the fact that it takes a long time to do it (i.e. the propagator $G_t$ is non-integrable). From this point of view, anomalous impact stems from market efficiency and from the presence of a long range correlated order flow.

\subsection{A Markovian  model for the latent order book}
\label{sec:MarkovBook}
The model described in the above section is an effective one, which does not take into account the fact that in a double auction market the price moves are determined by the local condition of the order book. The effect of a market order indeed depends upon the volume at the best, whereas limit orders and cancellations can change  prices even in absence of trades. Hence, if the description provided by the propagator model is at least qualitatively correct, it means that the kernel $G_{t-t^\prime}$ has to be thought of as an effective quantity incorporating a remarkable amount of information about the order book structure, rather than a fundamental property of the model. \\

A more accurate description of the price process, taking into account the one dimensional structure in which the price diffuses, can be formulated in the reference frame of the last execution price after trade number $t$ (which we denote by $\ell_t$). The best bid price just before trade $t+1$ is $b_t$ and the best ask price just before trade $t+1$ is $a_t$. Note that if the book does not evolve 
between trade number $t$ and just before trade number $t+1$, then either $a_t$ or $b_t$ is equal to $\ell_t$: the former case is when $\epsilon_t=+$ and the latter when $\epsilon_t=-$. If between the
two trades some activity has taken place in the book, then $\ell_t$ is in general not equal to either of them. With this convention, we show in appendix~\ref{app:MarkovBook} that for a Markovian 
evolution of the book one has:
\begin{equation}
\label{eq:ImpactMarkov}
   \langle \Delta \ell_t \rangle_\varphi = \langle \pi_{t} \rangle_\varphi + \varphi \langle s_{t} \rangle_\varphi  \qquad \Delta \ell_t := \ell_{t+1}-\ell_t,
\end{equation}
where the average $\langle \dots \rangle_\varphi$ is over all possible evolutions of the order book, and the mid-price $\pi_t$ and the half-spread $s_t$ are given by: 
\begin{eqnarray}
  \pi_t &=& \frac{a_t + b_t}{2} \\
  s_t &=& \frac{a_t - b_t}{2}.
\end{eqnarray}
Note that because prices are counted from $\ell_t$, $\pi_t$ tends to be positive after a sell and negative after a buy. 

Eq.~(\ref{eq:ImpactMarkov}) is analogous to Eq.~(\ref{eq:ImpactPropModel}) for the impact. It expresses the fact that under a constant bias the imbalance parameter $\varphi$ is linearly coupled to the spread, while the average asymmetry of the book is captured by the term $\pi_t$, which is zero by symmetry when $\varphi=0$. One expects that in the presence of a positive bias, trades are more 
likely to happen at the ask, and therefore $\langle \pi_{t} \rangle_\varphi$ is negative and partly compensates the second term. In order for the model to describe strictly anomalous impact, 
one should impose the ``slow absorption'' condition, analogous to the one of the previous section, which in this context would read:
\begin{equation}
\langle \pi_{t} \rangle_\varphi  + \varphi  \langle s_{t} \rangle_\varphi \sim t^{-\beta}\; ,
\end{equation}
with $\beta > 0$ describing the slow relaxation of the average price change towards zero:
\begin{equation}
\langle \pi_{\infty} \rangle_\varphi + \varphi \langle s_{\infty} \rangle_\varphi = 0 \;.
\end{equation}
These conditions express the fact that in order for the impact to be truly anomalous, the book should accumulate an asymmetry large enough to {\it exactly} compensate  the spread term which is linearly coupled to the bias. Otherwise, the impact must be linear in $\varphi$ and no asymptotic concavity can be present. The speed at which such asymmetry forms should then control the impact exponent: a quickly adapting 
book would give an integrable contribution ($\beta > 1$) to the average price change, and a bounded impact, whereas the presence of long memory in the order book would generate a non-trivial
impact exponent $\delta=1 - \beta$.

The diffusion properties of the model (corresponding to Eq.~(\ref{eq:DiffPropModel}) in the propagator model) can be obtained from the autocorrelation function:
\begin{widetext}
\begin{equation}
\label{eq:ACMarkov}
\langle \Delta \ell_t  \Delta \ell_0 \rangle_\varphi - \langle \Delta \ell_t \rangle_\varphi \langle \Delta \ell_0 \rangle_\varphi = s_0 \,(1-\varphi^2)\left( \frac{\langle \pi_t \rangle_{\varphi,+} - \langle \pi_t \rangle_{\varphi,-} }{2} \right)  
 + s_0 \,(1-\varphi)^2 \left(\frac{ \langle s_t \rangle_{\varphi,+} + \langle s_t \rangle_{\varphi,-}}{2} \right) g_t \nonumber
\end{equation}
\end{widetext}
where the symbol $\langle \cdot \rangle_{\varphi,\epsilon_0}$ indicates an average conditional to the sign of the first trade $\epsilon_0$ and $s_0$ denotes the stationary value of the half-spread. 
Again, the details about the derivation of these equations can be found in Appendix~\ref{app:MarkovBook}. Notice that these results are independent of the particular choice of dynamics for the book; the information about the time evolution of the book (or equivalently, the information about what happens in between market orders) is fully encoded in the averages $\langle \dots \rangle_\varphi$ and $\langle\dots\rangle_{\varphi,\epsilon_0}$.

\subsection{Comparison between the two models and open problems}

Unlike for the propagator model, we now have two degrees of freedom to determine the average price change: the bias $\varphi$ is coupled to the quantity  $\langle s_t \rangle$ (equivalently, each unbalanced buy trade pushes price up by half of the spread), while the mid-price $\langle \pi_t \rangle$ is independent of the bias, and accounts for the average pressure due to the book shape. In particular if more volume is available on the ask side, $\langle \pi_t \rangle $ is negative because large price changes are more likely to occur on the bid side, where the density of orders is smaller. This is consistent with what we have numerically shown in section~\ref{sec:Impact}: in our synthetic market a bias in the order flow is partially compensated, while the book slowly relaxes to a perturbed stationary value in which more volume sits on the ask side of the book. In the numerical model, the slow relaxation of the book takes place in the regime  $\mu^{-1} \ll \tau \ll \tau_\nu$, within
which an anomalous response can indeed build up.

Another important difference between the propagator model and the $\varepsilon$-intelligence model concerns the impact exponent. In the propagator model, imposing that prices are diffusive uniquely fixes the long time behavior of $G_{t-t^\prime}$ in terms of the correlation of the flow, as recalled above. This then immediately fixes the impact exponent as $\delta=(1+\gamma)/2$. This is the case because $G_{t-t^\prime}$ controls both the mean and  the variance of the price process. Under our general order book model, instead, we now deal with two independent quantities subject to two different boundary conditions ($\langle \pi_t \rangle_{\varphi,\epsilon_0}$ and $\langle s_t\rangle_{\varphi,\epsilon_0}$). This gives more flexibility to the model, which means that there is no longer a unique relation between the correlation exponent $\gamma$ and the impact exponent $\delta$, which will depend upon the details of the order book dynamics.

\begin{figure*}[htb]
  \centering
  \includegraphics{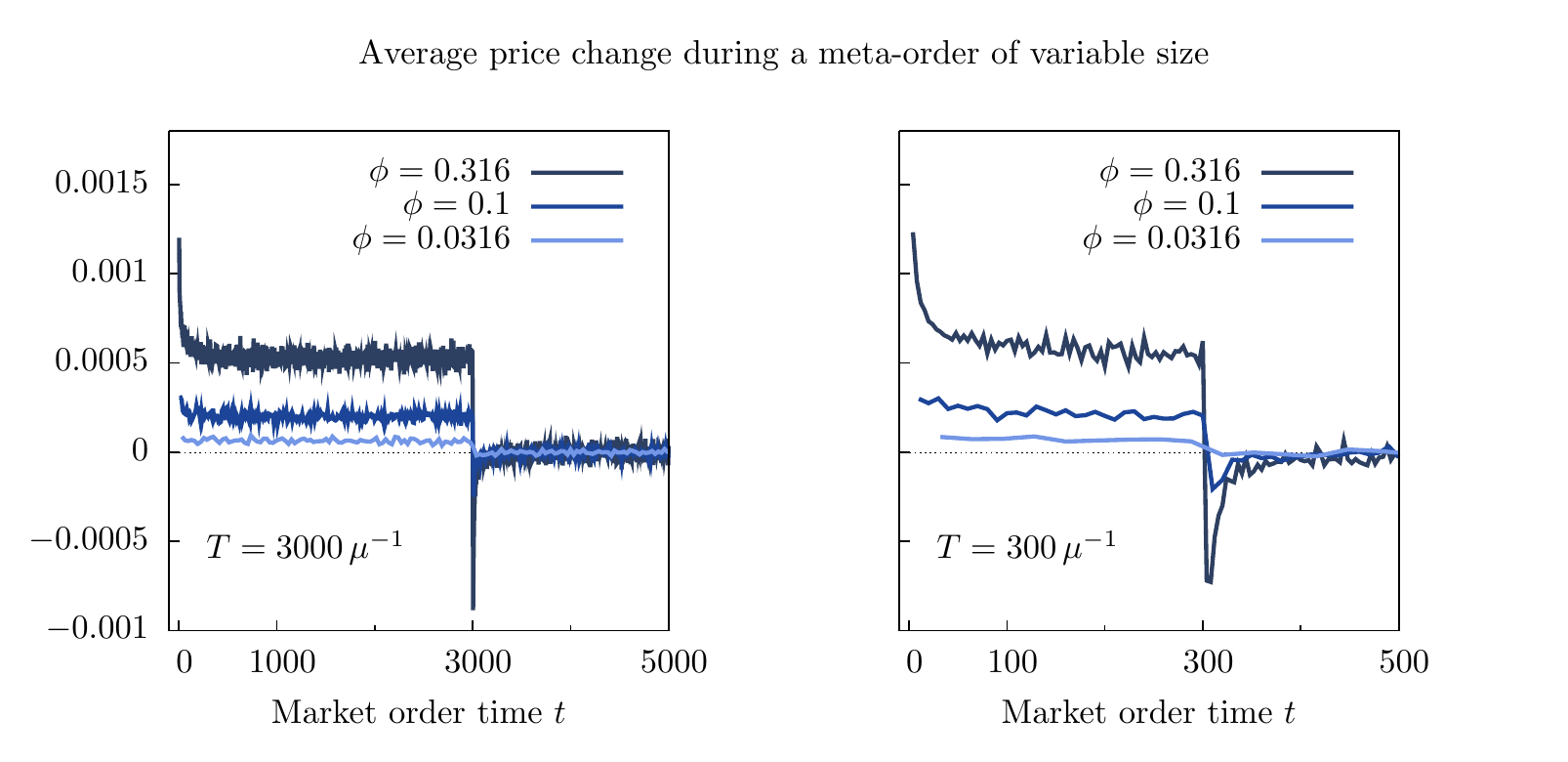}
  \caption{Average price change during and after the execution of a meta-order for the same choice of parameters as in Fig.~\ref{fig:RelaxationEpsSubmission} with $T=3000\, \mu^{-1}$ (left plot) and with $T=300\, \mu^{-1}$ (right plot) . The area under the curve represents the price impact of the meta-order. The figures show the transition to the asymptotic linear regime occurring for extremely long orders ($T\sim 1000 \, \mu^{-1}$), as opposed to the initial transient regime which causes concave impact. Also notice that the dynamics always tends to decrease the average price change, as it is expected in a market in which arbitrage opportunities are reduced as a consequancy of trading itself.}
  \label{fig:AvgPriceChange}
\end{figure*}

It is interesting to notice that within our framework the object manipulated in order to study market impact is the average price change $\langle \Delta \ell_t \rangle_\varphi$, rather than the impact itself. In this language the concavity of market impact corresponds to the fact that once a bias is added to the flow of orders, $\langle \Delta \ell_t \rangle_\varphi$ will progressively decrease such as to compensate (partially or completely) the bias. In other words, this point of view suggests the idea that the expected price change $\langle \Delta \ell_t \rangle_\varphi$ is a Lyapunouv function of the order book dynamics, as required by the assumption of market (statistical) efficiency: once some predictability (the bias in the order flow) is introduced, one expects that the dynamics of the market will remove arbitrage opportunities\footnote{This can be rigorously shown in other market models such as the Minority Game setting of~\cite{Barato:2013}. However in the setting of the Minority Game the expected price change is reduced by the action of market participants which by construction adapt in order to remove predictability from the market. In this context the reason for such decrease is purely mechanical, because sequence of orders on one side of the market tend to hit progressively higher volumes.}. Simulation results support this interpretation, as demonstrated by Fig.~\ref{fig:AvgPriceChange}, where we show that after a transient following the beginning of the meta-order, the average price change relaxes to a non-zero stationary value (as the impact has a residual permanent component). Once the meta-order is over, another transient follows and finally at large times the average price change reverts back to zero.
As a final note, we remark that the $\varepsilon$-intelligence market presented in section \ref{sec:Model} fulfills all the hypotheses required for the theoretical framework presented in this section to hold. In other words, we should in principle be able to provide an {\it exact} description of our synthetic market.\footnote{The fact that the $\varepsilon$-intelligence agents choose how much liquidity to take according to the volume on the best does not introduce a dependence on the book of the process $\epsilon_t$: although the volume taken is conditioned to the book shape, the sign of the orders is not. Another necessary condition for the validity of this description is the absence of \emph{trades through} (i.e., trades hitting multiple price levels), which are absent by construction in an $\varepsilon$-intelligence market. The only caveat which should be considered is that one needs $\zeta^\prime = \zeta$: the market orders associated with the meta-order and with the environment should consume liquidity in the same way if one wants to describe the modified flow through Eqs.~(\ref{eq:MarkOrderBias}) and (\ref{eq:MarkOrderCorr}).}
However, completing this program is technically difficult and goes beyond the scope of this paper. Whereas Eq.~(\ref{eq:ACMarkov}) relates the evolution of average price at time $t+1$ with the expected 
best quotes at time $t$, one would need to solve an infinite set of relations between the price gap at level $n$ and time $t+1$ with the gap at level $n+1$ and time $t$. This would correspond in continuous approximation to a partial differential equation for the dynamics of order book. We hope to come back to this problem in a future work.

\section{Conclusions}
\label{sec:Conclusions}

The aim of this work was to revisit and clarify the ``$\varepsilon$-intelligence'' model of T\'oth et al.~\cite{Toth:2011}, that was proposed as a minimal framework to understand the surprising non-additive, square-root dependence of the impact of meta-orders in financial markets. 
The basic mechanism, substantiated by the analytical and numerical results of~\cite{Toth:2011}, is that most of the daily liquidity 
is ``latent'' and furthermore vanishes linearly around the current price, as a consequence of the diffusion of the price itself, which depletes the nearby liquidity. Still, the numerical implementation of this
idea requires several extra specifications that are to some extent arbitrary, and the robustness of the scenario of T\'oth et al. needed to be ascertained. 

Our conclusions broadly support the universality of the results reported in~\cite{Toth:2011}, which is in itself important since the square-root dependence of the impact is empirically found to be independent of the market, epoch, microstructure, execution style, etc. It would be conceptually difficult to understand this universality if the theoretical results depended crucially on the microscopic specification of the model. This is in fact, in our opinion, one of the difficulty of the ``fair price'' scenario of Farmer et al.~\cite{Farmer:2011} which crucially depends, among other things, on the shape of the order-flow autocorrelation function (see for example Fig.~\ref{fig:GammaVSDelta}, which shows that the prediction $\delta=\gamma$ is not supported by our data on futures).

We have proposed and studied a variant of the model of T\'oth et al. where market efficiency does not rely entirely on the statistics of market orders (as in the original version~\cite{Toth:2011}),
but rather comes from the interaction between the flow of market order and the ``tit-for-tat'' reaction of limit orders, which tend to replenish the side of the book that is under pressure. This specification is far more realistic than the original one, and in fact allows us to account for the long-range correlation of the sign of both market and limit orders, an empirical fact that was not reproduced in~\cite{Toth:2011}. We find that the impact is even more concave in volume than in the original version of the model, and in fact closer to empirical results. 

We have also investigated different execution protocols, in particular one where the meta-order is executed {\it using limit orders only}, with the same qualitative result. This is an important finding, because there is a commonly held view that the impact of market orders is fundamentally different from that of limit orders. Although this is correct at the level of single trades~\cite{Eisler:2012}, empirical results suggest that the impact of meta-orders depend very little on the ratio of market to limit orders used for execution (see Fig.~\ref{fig:DISCUSImp}). 

Finally, we have shown that the transition from super-diffusion to sub-diffusion reported in \cite{Toth:2011} is in fact a cross-over that depends on the time scale over which the diffusion behaviour is 
probed. Although not hugely relevant for practical applications, this issue is of some importance from a theoretical point of view, as our model is an example of a random walk in an adaptive environment, for which very few mathematical results are available. We have shown how the original model can be slightly altered in order to give rise (at least numerically) to a genuine phase transition between  a super-diffusive and a sub-diffusive phase, such that purely diffusive motion is only realized on a co-dimension one sub-space of the parameters. It would be very interesting to obtain analytical results on this transition, on the time-dependent shape of the latent order book and on the impact of meta-orders within these simple models of markets. We have provided in the last section of this paper some ingredients that may enable one to achieve this program, within the framework of Markovian latent order books. As a general result, we have shown that anomalous, non-additive behaviour of impact requires that the liquidity buffer adapts slowly to the order flow~\cite{Barato:2013}, in such a way that the asymptotic price change induced by a meta-order vanishes. This suggests that expected price change acts as a Lyapunouv function of the order book dynamics, and as such is deeply related to market efficiency: once some predictability (the bias in the order flow) is introduced, one should expect that the dynamics of the market will act to remove arbitrage opportunities. At least conceptually, this is close to the arguments put forth in~\cite{Farmer:2011, Donier:2012}, although both the detailed ingredients and the conclusions differ in the two approaches. These ideas, and their precise relation with the ``microstructure invariance'' of Kyle \& Obizhaeva \cite{Kyle:2012}, would be well worth elucidating further.

By and large, our study lends strong support to the idea that the square-root impact law is a very generic and robust property, and requires very few ingredients to be valid. It is expected to hold in any market, provided the correlation time of the latent liquidity is much longer than the inter-transaction time. We believe that the impact on the implied volatility of option markets, for example, will
show a similar concavity. This would support the use of this square-root impact law to discount the expected cost of liquidation from the mark-to-market value of positions, as advocated in~\cite{Caccioli:2012}. 

\begin{acknowledgments}
We warmly thank our collaborators J. De Lataillade, C. Deremble, Z. Eisler, J. Kockelkoren and Y. Lemp\'eri\`ere for many very useful discussions. We have also benefitted from 
interesting comments and suggestions by R. Benichou, X. Brokmann, J. Donier, D. Farmer, J. Gatheral, A. Kyle, C. Lehalle, F. Lillo, M. Potters and H. Waelbroeck.  
\end{acknowledgments}

\appendix

\section{The Markovian book model:  mean and correlations}
\label{app:MarkovBook}
In section \ref{sec:MarkovBook} we presented an order book model constructed in the moving reference frame of the last execution price $\ell_t$ (where $t$ labels trade time), and linked its properties to the ones of the sign process $\epsilon_t$. Here we want to show that Eqs.~(\ref{eq:ImpactMarkov}) and (\ref{eq:ACMarkov}) describing the evolution of the last execution price can be derived by using basic properties of a Markovian order book, which is defined as the one in which the state right before the trade $t+1$, denoted by $\rho^\epsilon_{t+1}$, depends just upon the state before trade number $t$, $\rho^\epsilon_{t}$, and upon the sign of the $t$-th trade $\epsilon_t$.
This condition can equivalently be written as
\begin{equation}
  \label{eq:DefMarkovBook}
  p(\rho^\epsilon_{t+1} | \rho^\epsilon_0,\dots, \rho^\epsilon_t,\epsilon_1,\dots, \epsilon_t)= p(\rho^\epsilon_{t+1} | \rho^\epsilon_t, \epsilon_t) \; .
\end{equation}
In order to derive a master equation for $p(\rho^\epsilon_t | \rho^\epsilon_0, \epsilon_0)$, one needs to specify the statistics for the process $\epsilon_t$, which we assume to be defined by the relations
\begin{eqnarray}
  p(\epsilon_t) &=& \frac{1+\varphi \epsilon_t}{2} \label{eq:UncondSignAvg} \\
  p(\epsilon_t,\epsilon_0) &=& \varphi^2 \left( \frac{1+ \epsilon_t}{2} \right) \left( \frac{1+ \epsilon_0}{2} \right)\\ \nonumber
&+& \varphi (1-\varphi)   \left( \frac{1+ \epsilon_t}{2} \right) \left( \frac12 \right)\\ \nonumber
&+& (1-\varphi)\varphi  \left( \frac12 \right) \left( \frac{1+ \epsilon_0}{2} \right)\\ \nonumber
&+& (1-\varphi)^2 \left( \frac{1+g_t\epsilon_t\epsilon_0}{4} \right) \; . \label{eq:CondSignAvg}
\end{eqnarray}
This describes a market in which a fraction $\varphi$ of traders is submitting orders of fixed, positive sign, while the remaining $1-\varphi$ are correlated among themselves (although uncorrelated with the buyers). These are exactly the conditions specifying a meta-order submission process under the $\varepsilon$-intelligence model (as described in section \ref{sec:Impact}).
In this regime, one can derive a master equation for the evolution of the book via
\begin{widetext}
\begin{eqnarray}
  \label{eq:MasterEqProb}
  p(\rho^\epsilon_{t+1}| \rho^\epsilon_0,\epsilon_0) &=& \sum_{\rho^\epsilon_t,\epsilon_t} p(\rho^\epsilon_t| \rho^\epsilon_0,\epsilon_0) p(\rho^\epsilon_{t+1} | \rho^\epsilon_t,\epsilon_t) p(\epsilon_t | \epsilon_0) \\ \label{eq:MasterEqProbInteg}
&=& \sum_{\rho^\epsilon_t} p(\rho^\epsilon_t| \rho^\epsilon_0,\epsilon_0) \left[ \frac{p(\rho^\epsilon_{t+1} | \rho^\epsilon_t, + )+p(\rho^\epsilon_{t+1} | \rho^\epsilon_t, -)}{2} 
+  \frac{p(\rho^\epsilon_{t+1} | \rho^\epsilon_t, + ) - p(\rho^\epsilon_{t+1} | \rho^\epsilon_t, -)}{2} \left( \varphi + g_t \frac{(1-\varphi)^2\epsilon_0}{1+\varphi \epsilon_0} \right) \right] \; , \nonumber
\end{eqnarray}
\end{widetext}
which is an evolution equation for the probability of observing a book in a specific configuration at a given instant of time, given a starting condition $\rho^\epsilon_0$ and a sign $\epsilon_0$ for the first trade.
The master equation for the unconditional probability $p(\rho^\epsilon_{t+1}| \rho^\epsilon_0)$ can be analogously obtained (either by summing Eq.~(\ref{eq:MasterEqProbInteg}) by $\epsilon_0$ with the appropriate weights $p(\epsilon_0)$, or by directly using Eq.~(\ref{eq:MasterEqProb}) in order to obtain the evolution equation). In any case, the equation for the unconditional evolution corresponds to~(\ref{eq:MasterEqProbInteg}) with the substitution $g_t = 0$.
Notice that in Eq.~(\ref{eq:MasterEqProbInteg}) the effect of market orders has been integrated out, and is kept into account through the terms proportional to $\varphi$ and $g_t$. Interestingly, even though the market order process can have long memory,  Eq.~(\ref{eq:MasterEqProbInteg}) only couples subsequent times. \\ \\
By defining the mid-price $\pi_t$ and the half-spread $s_t$, one can use Eq.~(\ref{eq:MasterEqProbInteg}) to calculate the evolution of their averages. In particular Eq.~(\ref{eq:ImpactMarkov}) can be obtained by multiplying the master equation for the unconditional average by $\Delta \ell_t$ and summing over the book configurations $\rho^\epsilon_{t+1}$. In order to finally recover Eq.~(\ref{eq:ImpactMarkov}), one has then to use the fact that (in absence of trade through) it holds
\begin{eqnarray}
  \label{eq:EqPriceChangeBid}
  \sum_{\rho^\epsilon_{t+1}} \Delta \ell_t \,p(\rho^\epsilon_{t+1}| \rho^\epsilon_t ,+)& = & a(\rho^+_t) = a_t \\
  \sum_{\rho^\epsilon_{t+1}} \Delta \ell_t \,p(\rho^\epsilon_{t+1}| \rho^\epsilon_t ,-)& = & b(\rho^-_t) = b_t \; ,
\label{eq:EqPriceChangeaAsk}
\end{eqnarray}
where $a(\rho^+_t)$ and $b(\rho^-_t)$ are functions of the book state $\rho^\epsilon$ expressing the ask and the bid price.
Notice that a relation analogous to (\ref{eq:ImpactMarkov}) can be derived in terms of  conditional averages, so that it is also possible to write
\begin{eqnarray*}
  \label{eq:AltMarkovImpact}
  \langle \Delta \ell_t \rangle_\varphi &=& \left( \frac{\langle \pi_{t} \rangle_{\varphi,+} + \langle \pi_{t} \rangle_{\varphi,-}}{2} \right) \\ 
 &+& \varphi \left( \frac{\langle \pi_{t} \rangle_{\varphi,+} - \langle \pi_{t} \rangle_{\varphi,-}}{2} + \frac{\langle s_{t} \rangle_{\varphi,+} + \langle s_{t} \rangle_{\varphi,-}}{2}  \right) \; , \nonumber
\end{eqnarray*}
which relates the conditional and unconditional values of the averages. In particular by matching term by term Eqs.~(\ref{eq:ImpactMarkov}) and~(\ref{eq:AltMarkovImpact}) one obtains
\begin{eqnarray}
\langle \pi_{t} \rangle_{\varphi}  &=& \frac{\langle \pi_{t} \rangle_{\varphi,+} + \langle \pi_{t} \rangle_{\varphi,-}}{2} \\
\langle s_{t} \rangle_{\varphi,\epsilon_0} &=& \langle s_{t} \rangle_{\varphi,-\epsilon_0} \\
\langle \pi_{t} \rangle_\varphi - \langle \pi_{t} \rangle_{\varphi,\epsilon_0} &=& -\epsilon_0 \left( \langle s_{t} \rangle_\varphi - \langle s_{t} \rangle_{\varphi,\epsilon_0} \right) \; .
\end{eqnarray}
The expression of the autocorrelation function for this process is slightly more involved to derive: one can obtain Eq.~(\ref{eq:ACMarkov}) by considering a symmetric initial condition for the book and by writing
\begin{equation}
  \langle \Delta \ell_t  \Delta \ell_0 \rangle_\varphi = s_0\sum_{\epsilon_0} \epsilon_0  \left( \frac{1 + \varphi \, \epsilon_0}{2} \right) \langle \Delta \ell_t  \rangle_{\varphi,\epsilon_0} \; .
\end{equation}
After using the master equation (\ref{eq:MasterEqProbInteg}) in order to obtain the relation
\begin{equation}
  \langle \Delta \ell_t  \rangle_{\varphi,\epsilon_0} = \langle \pi_{t} \rangle_{\varphi,\epsilon_0} + \left( \varphi + g_t \frac{(1-\varphi)^2 \epsilon_0}{1+\varphi \epsilon_0} \right) \langle s_t \rangle_{\varphi,\epsilon_0} \, ,
\end{equation}
Eq.~(\ref{eq:ACMarkov}) can then be recovered.


\begin{thebibliography}{9}

  \bibitem{Kyle:1985}
  Kyle, A. (1985). {\it Continuous auctions and insider trading}. Econometrica: Journal of the Econometric Society, 1315-1335.

  \bibitem{Toth:2011}
  Toth, B., Lemp\'eri\`ere, Y., Deremble, C., De Lataillade, J., Kockelkoren, J., Bouchaud, J. P. (2011).
  {\it Anomalous price impact and the critical nature of liquidity in financial markets}. Physical Review X, 1(2), 021006.

  \bibitem{Torre:1997}
  Torre, N. (1997). {\it BARRA Market Impact Model Handbook}. BARRA Inc., Berkeley.

  \bibitem{Almgren:2005}
  Almgren, R., Thum, C., Hauptmann, E., Li, H. (2005). {\it Direct estimation of equity market impact}, Risk, 18(7), 5762.

  \bibitem{Moro:2009}
  Moro, E., Vicente, J., Moyano, L. G., Gerig, A., Farmer, J. D., Vaglica, G., Lillo, F., Mantegna, R. N. (2009).{\it Market impact and trading profile of hidden orders in stock markets.} Physical Review E, 80(6), 066102.

  \bibitem{Kyle:2012}
  Kyle, A. P., Obizhaeva, A. (2012). {\it Large Bets and Stock Market Crashes.} Available at SSRN 2023776.

  \bibitem{Bershova:2013}
  Bershova, N., Rakhlin, D. (2013). {\it The non-linear market impact of large trades: Evidence from buy-side order flow.} Available at SSRN 2197534.

  \bibitem{Waelbroeck:2013}
  Waelbroeck, H., Gomes, C. (2013). {\it Is Market Impact a Measure of the Information Value of Trades? Market Response to Liquidity vs. Informed Trades.} Available at SSRN 2291720.

\bibitem{Farmer:2011}
  Farmer, J. D., Gerig, A., Lillo, F., and Waelbroeck, H. (2011). {\it How efficiency shapes market impact}. arXiv preprint arXiv:1102.5457.
  
  \bibitem{Bak:1997}
  Bak, P., Paczuski, M., Shubik, M. (1997). {\it Price variations in a stock market with many agents}. Physica A: Statistical Mechanics and its Applications, 246(3), 430-453.

  \bibitem{Tang:1999}
  Tang, L. H., Tian, G. S. (1999). {\it Reaction-diffusion-branching models of stock price fluctuations}. Physica A: Statistical Mechanics and its Applications, 264(3), 543-550.


\bibitem{Bouchaud:2006}
Bouchaud, J. P., Kockelkoren, J., Potters, M. (2006). {\it Random walks, liquidity molasses and critical response in financial markets.} Quantitative Finance, 6(02), 115-123.

  \bibitem{Weber:2005}
Weber, P., and Rosenow, B. (2005). {\it Order book approach to price impact}. Quantitative Finance, 5(4), 357-364.

  \bibitem{Bouchaud:2008}
Bouchaud, J. P., Farmer, J., Lillo, F. (2008). {\it How markets slowly digest changes in supply and demand}, in Handbook of Financial Markets: Dynamics and Evolution, pp 57-156. Eds.
Thorsten Hens and Klaus Schenk-Hoppe. Elsevier: Academic Press, (2008)

  \bibitem{Sandas:2001}
  Sandas, P. (2001). {\it Adverse selection and comparative market making:empirical evidence
from a limit order market}. The Review of Financial Studies, 14(3),705-734.


  \bibitem{Farmer:2003}
Smith, E., Farmer, J. D., Gillemot, L., Krishnamurthy, S. (2003). {\it Statistical theory of the continuous double auction.} Quantitative Finance, 3(6), 481-514.

  \bibitem{Farmer:2005}
Farmer, J. D., Patelli, P., and Zovko, I. I. (2005). {\it The predictive power of zero intelligence in financial markets}. Proceedings of the National Academy of Sciences of the United States of America, 102(6), 2254-2259.

\bibitem{Hawkes:recent} For recent work on activity clustering within a Hawkes process description, see e.g Bacry,~E., Dayri,~K., and Muzy,~J.-F. (2012). {\it Non-parametric kernel estimation for symmetric Hawkes processes. Application to high frequency financial data.} European Journal of Physics  B 85(5), 157, and Refs.\ therein.

\bibitem{Lillo:2005}
Lillo, F., Mike, S., and Farmer, J. D. (2005). {\it Theory for long memory in supply and demand.} Physical Review E, 71(6), 066122.

\bibitem{Bouchaud:2004}
Bouchaud, J. P., Gefen, Y., Potters, M., and Wyart, M. (2004). {\it Fluctuations and response in financial markets: the subtle nature of ``random'' price changes.} Quantitative Finance, 4(2), 176-190.

  \bibitem{Lillo:2004}
  Lillo, F., and Farmer, J. D. (2004). {\it The long memory of the efficient market}. Studies in Nonlinear Dynamics and Econometrics, 8(3).

  \bibitem{Wyart:2008}
  Wyart, M., Bouchaud, J. P., Kockelkoren, J., Potters, M., Vettorazzo, M. (2008).{\it  Relation between bid-ask spread, impact and volatility in order-driven markets. }Quantitative Finance, 8(1), 41-57.

  \bibitem{Maslov:2000}
  Maslov, S. (2000). {\it Simple model of a limit order-driven market}. Physica A: Statistical Mechanics and its Applications, 278(3), 571-578.


  \bibitem{Cornell:1995}
  Cornell, S. J. (1995). {\it Refined simulations of the reaction front for diffusion-limited two-species annihilation in one dimension.} Physical Review E, 51(5), 4055.


  \bibitem{Cardy:1996}
  Barkema, G. T., Howard, M. J., Cardy, J. L. (1996). {\it Reaction-diffusion front for A+B $\to \emptyset$ in one dimension.} Physical Review E, 53(3), R2017.


  \bibitem{Eliezer:1998}
  Eliezer, D., Kogan, I. I. (1998). {\it Scaling laws for the market microstructure of the interdealer broker markets}. arXiv preprint cond-mat/9808240, unpublished.

  
  
  \bibitem{Donier:2012} Donier, J. {\it Market Impact with Autocorrelated Order Flow under Perfect Competition}. arXiv:1212.4770

  \bibitem{Bouchaud:2002}
  Bouchaud, J. P., M\'ezard, M., Potters, M. (2002). {\it Statistical properties of stock order books: empirical results and models}. Quantitative Finance, 2(4), 251-256.

  \bibitem{Biais:1995}
  Biais, B., Hillion, P., and Spatt, C. (1995). {\it An empirical analysis of the limit order book and the order flow in the Paris Bourse}. Journal of Finance, 50(5), 1655-1689.

  \bibitem{Eisler:2012}
  Eisler, Z., Bouchaud, J. P., Kockelkoren, J. (2012). {\it The price impact of order book events: market orders, limit orders and cancellations}. Quantitative Finance, 12(9), 1395-1419.

  \bibitem{Barato:2013}
  Barato, A. C., Mastromatteo, I., Bardoscia, M., and Marsili, M. (2013).{\it  Impact of meta-order in the Minority Game}. Quantitative Finance, 13(9), 1343-1352.
  
  \bibitem{Caccioli:2012} Caccioli, F., Bouchaud, J. P., Farmer, J. D. (2012). {\it Impact-adjusted mark-to-market valuation and the criticality of leverage}. 
  arXiv preprint 1204.0922; Risk Magazine, May 2012;

 

\end{thebibliography}
\end{document}